\documentclass[twocolumn,amsmath,amsfonts,amssymb,aps,prx,preprintnumbers,superscriptaddress]{revtex4-2}

\usepackage[utf8]{inputenc}
\usepackage[T1]{fontenc}
\usepackage{lmodern}
\usepackage{graphicx}
\usepackage{dcolumn}
\usepackage{bm}
\usepackage{textcomp}
\usepackage{ulem}

\usepackage{ifpdf}
\usepackage{physics}
\usepackage{float}
\usepackage[squaren,Gray]{SIunits}
\usepackage{color}
\definecolor{red}{rgb}{1,0,0}
\definecolor{blue}{rgb}{0,0,1}
\definecolor{darkred}{rgb}{0.6,0,0}
\definecolor{darkblue}{rgb}{0,0,0.6}
\definecolor{darkgreen}{rgb}{0,0.5,0}
\definecolor{grey}{rgb}{0.5,0.5,0.5}
\definecolor{black}{rgb}{0,0,0}

\bibstyle{chem-acs}

\ifpdf
\usepackage{epstopdf}
\usepackage[pdftex,unicode,pdfstartview={FitH},pdfborder={0 0 0}]{hyperref}
\usepackage{hypcap}
\else
\usepackage[hypertex]{hyperref}
\fi
\hypersetup{
    bookmarksnumbered = true,
}

\newcolumntype{R}{>{$\displaystyle}r<{$}}
\newcolumntype{C}{>{$\displaystyle}c<{$}}

\begin{document}

\title{Excitonic fingerprints of magnetic configurations and switching in multilayer CrSBr}

\author{Lukas Krelle}
\affiliation{Institute for Condensed Matter Physics, TU Darmstadt, Hochschulstraße 6-8, D-64289 Darmstadt, Germany}

\author{Ryan Tan}
\affiliation{Institute for Condensed Matter Physics, TU Darmstadt, Hochschulstraße 6-8, D-64289 Darmstadt, Germany}

\author{Jakob Conradi}
\affiliation{Institute for Condensed Matter Physics, TU Darmstadt, Hochschulstraße 6-8, D-64289 Darmstadt, Germany}


\author{Priyanka Mondal}
\affiliation{Institute for Condensed Matter Physics, TU Darmstadt, Hochschulstraße 6-8, D-64289 Darmstadt, Germany}

\author{Wenze Lan}
\affiliation{Institute for Condensed Matter Physics, TU Darmstadt, Hochschulstraße 6-8, D-64289 Darmstadt, Germany}

\author{Kseniia Mosina}
\affiliation{Department of Inorganic Chemistry, University of Chemistry and Technology Prague, Technicka 5, 166 28 Prague 6, Czech Republic}


\author{Regine von Klitzing}
\affiliation{Institute for Condensed Matter Physics, TU Darmstadt, Hochschulstraße 6-8, D-64289 Darmstadt, Germany}



\author{Zdenek Sofer}
\affiliation{Department of Inorganic Chemistry, University of Chemistry and Technology Prague, Technicka 5, 166 28 Prague 6, Czech Republic}

\author{Bernhard Urbaszek}

\affiliation{Institute for Condensed Matter Physics, TU Darmstadt, Hochschulstraße 6-8, D-64289 Darmstadt, Germany}

\begin{abstract}

\textbf{ABSTRACT:} The coupling between electronic states and magnetism provides a route towards optical readout and control of magnetic information. In the magnetic semiconductor CrSBr, excitons are coupled to magnetic order, making their optical response sensitive to the underlying magnetization. Here, we show that the energy and oscillator strength of bulk and surface excitons provide distinct spectroscopic fingerprints of magnetic configurations and switching pathways. We distinguish domain-wall-mediated magnetization reversal, manifested by continuous spectral evolution as a domain wall traverses the optical spot, from abrupt, large-area magnetization reversal. Using the resulting excitonic fingerprints, we reconstruct successive magnetic configurations in 4- and 5-layer CrSBr during the transition from ferromagnetic to antiferromagnetic order. We further find that the sensitivity to magnetic order is strongly exciton-dependent: low-energy excitons resolve intermediate and surface-related configurations, whereas a higher-energy exciton predominantly exhibits a transfer of oscillator strength between ferromagnetic and antiferromagnetic resonances. These results establish excitonic spectroscopy as a sensitive probe of layer-dependent magnetic configurations and their switching pathways in layered magnetic semiconductors. \\ \\
\textbf{KEYWORDS:} CrSBr, magneto-optics, magnetic semiconductor, van der Waals materials, excitons, layered antiferromagnet

\end{abstract}

\maketitle

\section{Introduction}
\indent
Magnetic van der Waals materials offer promising platforms for spintronic and optoelectronic applications, owing to their low switching fields and the strong coupling between magnetic order and electronic and collective excitations, including excitons, phonons, magnons, and photons \cite{ahn20202d, mi2023two, adak2026excitons, park20262d}. Among these materials, CrSBr has emerged as a particularly versatile platform for exploring the interplay between magnetism and optical excitations, with reported magnon–exciton \cite{diederich2023tunable, datta2025magnon, bae2022exciton}, exciton–phonon \cite{lin2024strong}, and exciton–photon coupling \cite{dirnberger2023magneto, wang2023magnetically, budak2026role}. Its magnetic and optical properties can further be tuned through external electric and magnetic fields \cite{li2026electric, wilson2021interlayer, krelle2025magnetic}, electrostatic doping \cite{tabataba2024doping, graham2026space}, and materials-engineering approaches including interlayer twisting \cite{mondal2026twist, chen2024twist}, alloying \cite{badola2026van, smiertka2026tunable}, and ion irradiation \cite{long2023ferromagnetic, long2024rise, markina2026detecting}. Together, these properties make CrSBr an attractive platform for investigating and controlling the interplay between magnetism and optical excitations.

The strong sensitivity of excitons to the magnetic order in CrSBr provides a particularly promising route for optical readout of magnetization \cite{wilson2021interlayer, tabataba2024doping}. Changes in exciton energy, linewidth, and oscillator strength encode information about the underlying magnetic state, potentially allowing the magnetization of multiple layers to be accessed simultaneously within a single optical measurement \cite{lee2021magnetic}. This provides a complementary approach to local magnetic probes such as NV magnetometry, whose sensitivity is primarily determined by the magnetic fields emerging near the sample surface \cite{tschudin2024imaging, bagani2024imaging}. However, the microscopic relationship between the magnetic configuration and the excitonic response remains incompletely understood. This challenge becomes particularly pronounced in multilayer CrSBr, where the number of possible layer-dependent magnetic configurations increases rapidly with thickness \cite{lopion2025optical, sun2025resolving}. Distinguishing these configurations using conventional optical spectroscopy is further complicated by the finite optical spot size, which can spatially average over magnetic domains and domain walls. Recent identification of excitons associated with the surface and bulk layers of CrSBr \cite{shao2025magnetically, choi2026bulk} provides an additional spectroscopic handle for distinguishing the magnetic response of surface and interior layers.

Here, we investigate the sensitivity and limitations of excitonic spectroscopy as a probe of magnetization in multilayer CrSBr. We fabricate a staircase sample comprising 4–9 layers and perform cryogenic differential reflectance contrast ($\mathrm{DR/R}$) spectroscopy under applied magnetic fields. We distinguish two qualitatively different magnetization-switching processes: domain-wall-mediated reversal and abrupt large-area magnetization reversal, and identify their distinct spectroscopic signatures. Magnetic-field-dependent spatial line scans further allow us to directly correlate these optical signatures with the motion of domain walls and large-area switching events. Finally, by combining the energy and oscillator strength of selected bulk and surface exciton features with a simple model linking the optical response to the underlying magnetization, we infer the magnetic configurations of multilayer CrSBr and identify their characteristic optical fingerprints. Our results establish excitonic spectroscopy as a sensitive and spatially averaged approach for resolving magnetization reversal and magnetic-order transitions in layered magnetic semiconductors.
%

\section{Results and Discussion}


\begin{figure*}[t!]
\includegraphics[width = 17cm]{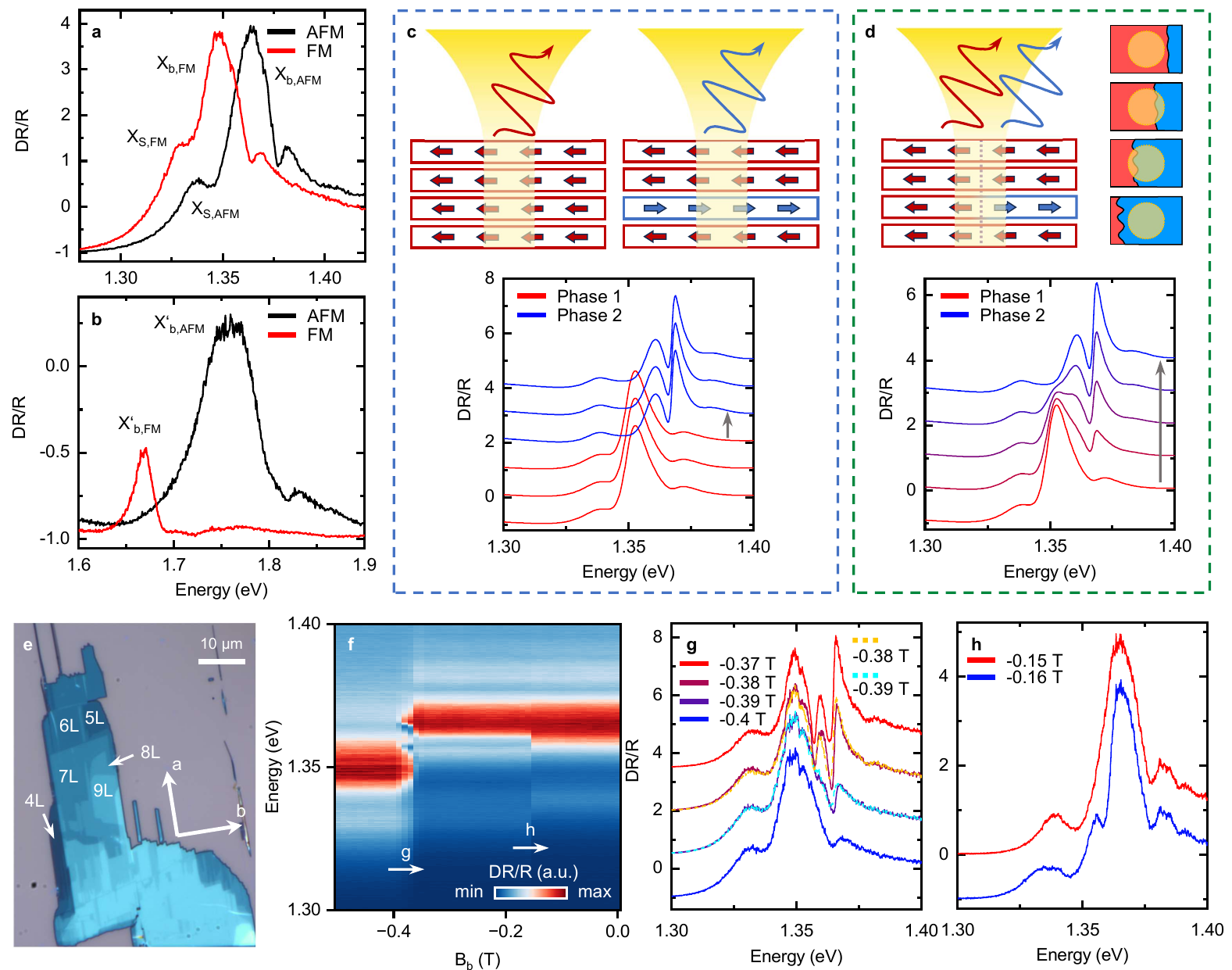}
\caption{\textbf{Domains and spin flips in spectroscopic mesurements.} \textbf{a, b} Exemplary DR/R of 9 layer CrSBr in the lower energy region (\textbf{a}) and the higher energy region (\textbf{b}). \textbf{c,d} Schematic representation of the influence of abrupt spin flips (\textbf{c}) and domain wall movement (\textbf{d}) on DR/R measurements using transfer-matrix simulations. \textbf{e} Microscope image of the exfoliated sample. \textbf{f} Magnetic field dependent DR/R measurement in the 6 layer region. Arrows indicate spectra shown in \textbf{g,h}. \textbf{g} DR/R signatures of a domain wall passing through the optical spot. Dotted graphs correspond to modelled spectra according to discussion. \textbf{h} DR/R signatures of an abrupt change in magnetization in the measurement of \textbf{f}} 
\label{fig:Domain_Flip_Sketch}
\end{figure*}


CrSBr is a layered A-type antiferromagnet, in which the spins sitting on Cr sites align ferromagnetically along the crystal b-axis within a layer and antiferromagnetically between adjacent layers. As a two-dimensional semiconductor it hosts tightly bound and highly optically active excitons. Several different types of exciton species have been identified in spectroscopic experiments at low temperatures \cite{wilson2021interlayer, tabataba2024doping, shao2025magnetically, smiertka2026distinct}. Around 1.36 eV and 1.34 eV CrSBr exhibits resonances stemming from excitons confined in bulk $X_b$ and surface layers $X_S$ respectively due to the opposite spin orientation in adjacent layers \cite{shao2025magnetically, choi2026bulk}. Due to their differences in magnetic and dielectric environment, the two resonances are shifted by $\approx$ 20-25 meV with respect to each other. Fig. \ref{fig:Domain_Flip_Sketch}a diplays an exemplary differential reflectance contrast (DR/R) measurement in the respective energy range, performed at $T = 4.7$ K in a confocal microscope \cite{shree2021guide} with an optical detection spot diameter of the order of $\lesssim 1~\mu$m. Recently, measurements in strong magnetic fields have attributed a more Frenkel-like character to these excitons \cite{smiertka2026distinct}. Additionally, the work identified another excitonic resonance with more Wannier-like character and larger exciton wavefunction around 1.77 eV, which will be referred to as $X_b'$ \cite{smiertka2026distinct, komar2024colossal}. Fig. \ref{fig:Domain_Flip_Sketch}b displays an exemplary DR/R measurement in the respective energy range showing a weaker resonance compared to $X_b$. All of the identified excitonic resonances exhibit an intricate interplay of their respective energies, widths and oscillator strengths with the magnetic order of the system. 


\begin{figure*}[t!]
\includegraphics[width = 17cm]{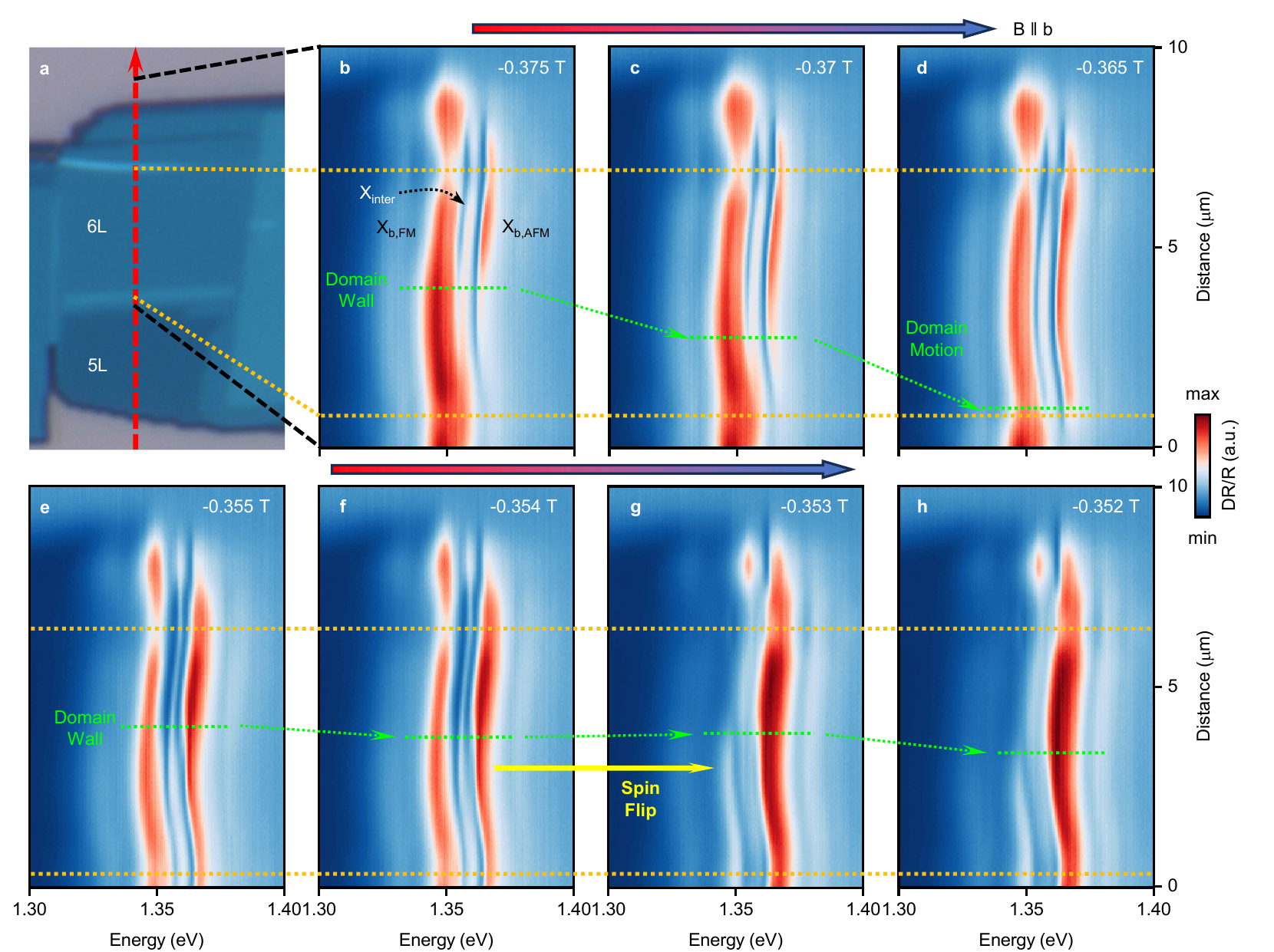}
\caption{\textbf{Domains and spin flips in linescans.} \textbf{a} Microscope image of the sample region across which linescans indicated by the red arrow were performed. \textbf{b-d} DR/R linescans showing the movement of a domain wall through the optical spot. Dotted green line indicates the domain wall and its movement in the spectra. Dotted black and yellow lines indicate shown regions of the linescan and sample features. \textbf{e-h} DR/R linescans showing a sudden flip of magnetization in the system without passage of a domain wall. Yellow arrow indicates the transition field for the magnetization flip.} 
\label{fig:Domain_Linescans_6L}
\end{figure*}


Due to the antiferromagnetic spin alignment between adjacent layers, excitons are confined to a single layer and tunneling to adjacent layers is forbidden, as long as no magnetic field is applied. By applying a magnetic field, the spins can be forced to align with the magnetic field direction driving the system from an antiferromagnetic (AFM) state to a ferromagnetic (FM) state after reaching the saturation field allowing charge transfer between layers \cite{wilson2021interlayer, ziebel2024crsbr}. If the magnetic field is applied along the crystal b-axis, the spins flip abruptly, while applying the field along the a- or c-axis, causes a canting of the spins. Due to changes in the bandstructure and exciton binding energy the exciton $X_b$ shifts around 15 meV to lower energies from AFM to FM state while $X_S$ shifts only about half as much, as reported in several studies \cite{wilson2021interlayer,tabataba2024doping,choi2026bulk}. In contrast the exciton at higher energies shows much larger shifts of about 90 meV \cite{smiertka2026distinct, komar2024colossal}. Fig. \ref{fig:Domain_Flip_Sketch}a and b show DR/R spectra in the FM configuration in the lower and higher energy range respectively, highlighting the difference in energy shift. \\

The coupling between excitons and magnetic order in CrSBr provides means of identifying different magnetic configurations through their optical response. In particular, changes in the energy and oscillator strength of excitonic resonances can encode the underlying layer-resolved magnetization, as different magnetic configurations give rise to distinct optical spectra. However, interpreting changes in the optical response requires careful consideration of the spatial averaging inherent to the finite optical spot. As illustrated in Fig. \ref{fig:Domain_Flip_Sketch}c,d, a spectral change can arise either from a change in the magnetic configuration that is homogeneous across the optical spot or from the coexistence of different magnetic configurations within the spot.\\
In the first case, the magnetic configuration changes uniformly over the illuminated area, corresponding to a large-area magnetization flip. The resulting optical response therefore changes abruptly within a narrow magnetic-field interval, reflecting the direct transition from one magnetic configuration to another. In the second case, the optical spot contains two magnetic domains separated by a domain wall, as illustrated in Fig. \ref{fig:Domain_Flip_Sketch}d. As the magnetic field drives the domain wall across the spot, the relative fractions of the two configurations within the probed area change continuously. The measured spectrum is consequently a spatially averaged combination of the optical responses of the two domains, leading to a gradual evolution of the spectral features rather than the abrupt change expected for a homogeneous magnetization flip.\\
This distinction is essential for interpreting the spectroscopic response: abrupt spectral changes indicate a large-area magnetic flip, whereas continuous changes can arise from domain-wall motion through the optical probe volume.


\begin{figure*}[t!]
\includegraphics[width = 15cm]{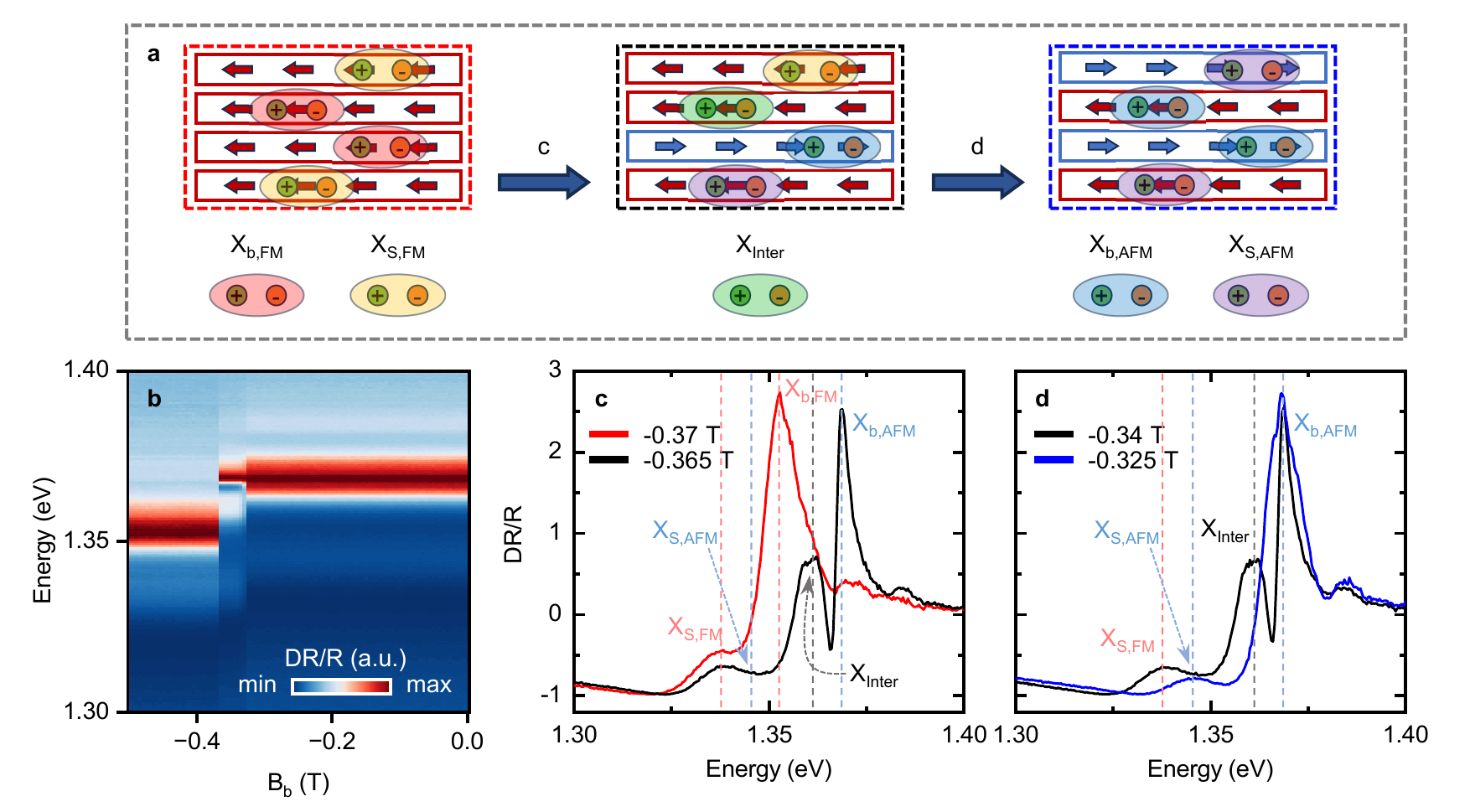}
\caption{\textbf{Magnetic configuration evolution for 4 layer CrSBr.} \textbf{a} Schematic of the magnetization configurations and excitons involved in the optical response. Arrows denote the transitions for the spectra shown in \textbf{c,d}. \textbf{b} Magnetic field dependent DR/R measurement in the 4 layer region. \textbf{c} DR/R spectra of the first change in magnetization in \textbf{a}. Dashed red and blue lines indicate the energies of the exciton resonances of the FM and AFM configuration respectively. Dashed black line indicates the energy of the intermediate resonance. \textbf{d} DR/R spectra of the second change in magnetization indicated in \textbf{a}. Dashed lines as in \textbf{c}.}
\label{fig:4L_Logic}
\end{figure*}


We want to elucidate the spectrally different response using a 6 layer region of the sample shown in Fig. \ref{fig:Domain_Flip_Sketch}e (see Fig. S1 for a topographic analysis of the sample using atomic force microscopy). For this purpose we performed DR/R spectroscopy at T = 4.7 K focusing on the lower energy region and applied a magnetic field along the magnetic easy axis (b-axis) displayed in Fig. \ref{fig:Domain_Flip_Sketch}f. We start the magnetic field sweep in the FM phase at $B_b = -0.5$ T and increase the field in 10 mT steps. At $B_b = -0.4$ T, the system still remains in the fully FM phase and displays surface and bulk excitonic resonances $X_{S,FM}$ and $X_{b,FM}$ associated with the FM phase. Between $B_b = -0.4$ T and $B_b = -0.37$ T, we observe a gradual change of the spectral response shown in Fig. \ref{fig:Domain_Flip_Sketch}g, in which the bulk exciton $X_{b,AFM}$ associated with the AFM phase and an intermediate resonance $X_{inter}$ emerge while $X_{b,FM}$ loses oscillator strength. The intermediate resonance has been observed in recent work and we will try to elaborate its origin in this work \cite{lopion2025optical, krelle2025magnetic}. Due to the rather continuous change in the spectral features, we assume a domain wall propagating through the spot that separates two spatial regions with spectra similar to $B_b = -0.4$ T and $B_b = -0.37$ T respectively. The spectra at $B_b = -0.39$ T and $B_b = -0.38$ T, can nicely be reproduced by adding the start and end spectra using a weight $DR/R_{Tot}= a\cdot DR/R_{B = -0.4\mathrm{T}}+b\cdot DR/R_{B = -0.37 \mathrm{T}}$, with $a+b=1$, where $a,b$ denote the fraction of the respective phases of the optical spot. At $B_b = -0.39 $ T and $B_b = -0.38$ T using $a = 0.8,\:b=0.2$ and $a = 0.22,\:b=0.78$ yielded reasonable agreement with the experiment. Additionally, we observe a sudden change in DR/R at $B_b = -0.16$ T after which the spectrum remains constant, indicating a flip of the systems magnetization. Fig. \ref{fig:Domain_Flip_Sketch}h displays DR/R spectra before and after the flip.

To investigate if a domain wall is indeed the source of the changes in Fig. \ref{fig:Domain_Flip_Sketch}g, we performed spatial linescans along the b-axis across the 6 layers region in Fig. \ref{fig:Domain_Linescans_6L}a for different magnetic field strengths $B_b$. Figs. \ref{fig:Domain_Linescans_6L}b-d display exemplary linescans for $B_b=-0.375$ T to $B_b=-0.365$ T. For comparison, in Fig. S2 we plot the same linescans for -0.5 T and 0.0 T. In all linescans we observe slight shifts of the resonance energies across the sample, which we attribute sample inhomogeneities. These energy shifts allow us to identify specific sample features marked by yellow lines in Fig. \ref{fig:Domain_Linescans_6L} in the linescans. At $B_b=-0.375$ T, we observe a domain wall separating a region with spectra of the complete FM configuration from a region of an intermediate magnetization configuration as observed in Fig. \ref{fig:Domain_Flip_Sketch}g. This is revealed by the presence of all three resonances $X_{b,FM}$, $X_{inter}$ and $X_{b,AFM}$. With increasing magnetic field the domain wall propagates closer to the edge of the 6 layers region until the FM region has fully vanished and only the intermediate configuration remains. In Fig. S3, we plot spectra of the linescan across the domain wall for $B_b=-0.375$ T and find the same type of spectral changes as in Fig. \ref{fig:Domain_Flip_Sketch}g. This further supports our claim of a domain wall responsible for the observed spectral changes.

The nucleation and motion of domain walls depend sensitively on sample boundaries, defects, strain, and exchange bias \cite{wang2025configurable, tschudin2024imaging, pellet2025lateral, bagani2024imaging}. Consistent with this sensitivity, we observe substantially different domain-wall dynamics across the sample (Figs. S4 and S5). Our linescans provide high spatial resolution along the $b$-axis but limited information along the perpendicular $a$-axis. The domain-wall orientation can nevertheless be inferred from the lateral extent of the spectral transition relative to the optical spot size ($\lesssim 1~\mu$m). For example, the nearly simultaneous spectral changes across a $\sim15~\mu$m linescan in Fig. s4 indicate a domain wall extending predominantly along the $b$-axis and moving approximately perpendicular to the linescan. Because the wall resides in layers common to the adjacent 7- and 8-layer regions, it can traverse both thickness regions simultaneously. In contrast, the domain wall in Fig. S5 is nearly immobile over the applied-field range and separates a fully FM region from an almost fully AFM region, characterized by a strong $X_{b,AFM}$ and weak $X_{inter}$ resonance. Its transition region is comparable to the optical spot size, consistent with a wall extending predominantly along the $a$-axis. Together with the data in Fig. \ref{fig:Domain_Linescans_6L}, these observations indicate that the domain wall in our main dataset is oriented diagonally between the $a$- and $b$-axes in the region covered by the optical spot.

Most of the spectral changes observed in our measurements are consistent with domain-wall-mediated magnetization switching. However, a clear domain-wall signature is not always resolved, because the magnetic-field step size can exceed the field interval required for the wall to traverse the optical spot. As a result, the domain-wall assisted transition occurs between consecutive measurements resembling a sudden flip in magnetization. We also observe events that cannot be attributed to domain-wall motion. Figs. \ref{fig:Domain_Linescans_6L}e–g show consecutive linescans of the same 6-layer region between $B_b=-0.355$ T and $-0.352$ T, acquired in 1 mT steps. Although we observe a domain wall that remains nearly stationary, the spectra undergo a pronounced global change between $B_b=-0.354$ T and $-0.353$ T. The absence of corresponding domain-wall motion suggests that this event arises from an abrupt, large-area reversal of the magnetization of specific layers, occurring independently of an observable domain wall.


\begin{figure*}[t!]
\includegraphics[width = 17cm]{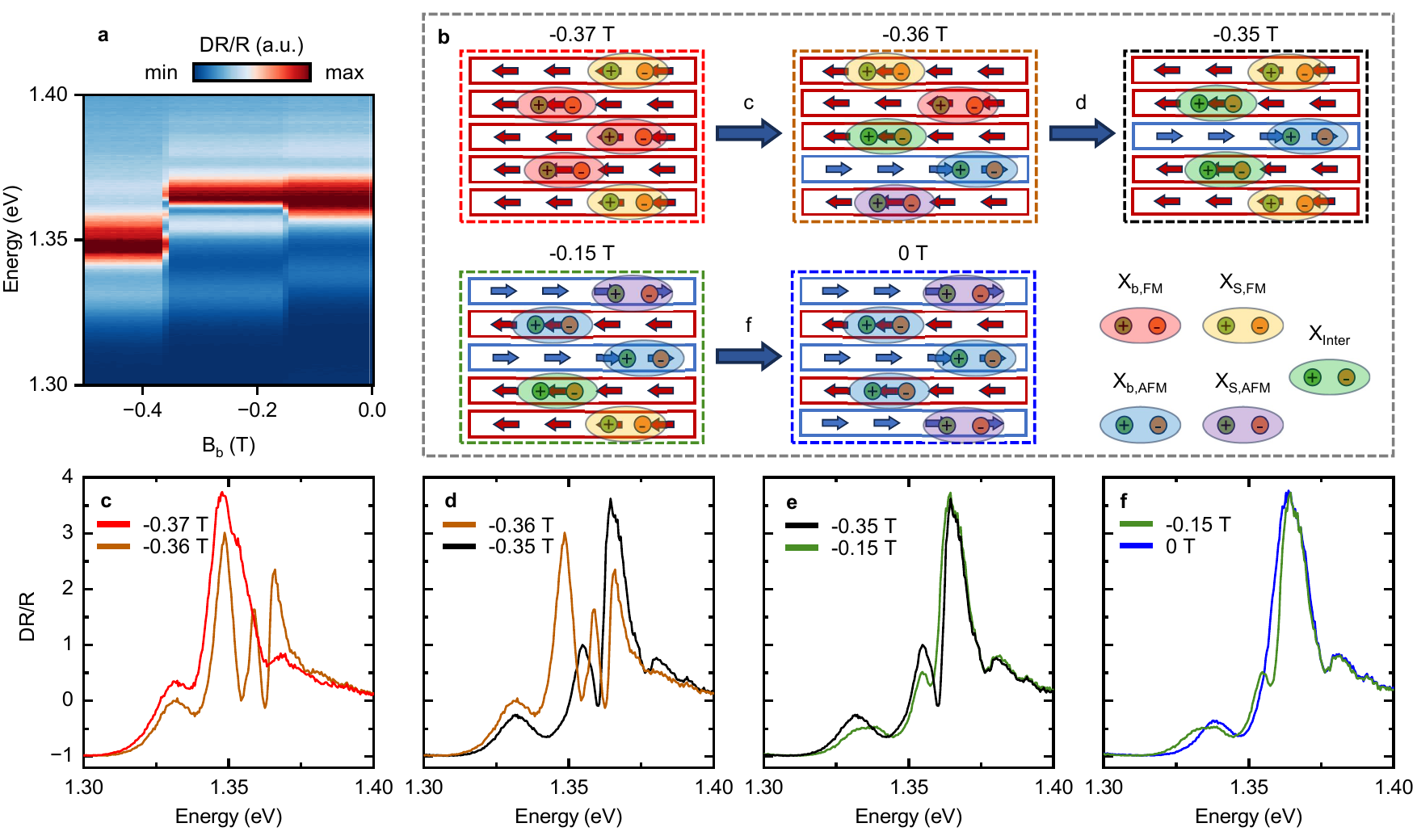}
\caption{\textbf{Magnetic configuration evolution for 5 layer CrSBr.} \textbf{a} Exemplary magnetic field dependent DR/R measurement in the 5 layer region. \textbf{b} Schematic of the magnetization configurations and excitons involved in the optical response. Arrows denote the transitions for the spectra shown in \textbf{c-f}. \textbf{c-f} DR/R spectra corresponding to the transitions proposed in \textbf{b}} 
\label{fig:5L_Logic}
\end{figure*}


We next use the spectral changes to infer the underlying magnetic configurations. As illustrated in Fig. \ref{fig:4L_Logic}a, we assume that the exciton energy is determined by the relative magnetization of adjacent layers. In bulk layers, antiparallel and parallel alignment correspond to $X_{b,AFM}$ and $X_{b,FM}$, respectively. For surface layers, the corresponding excitons are $X_{S,AFM}$ and $X_{S,FM}$, determined by the antiparallel or parallel alignment of the surface layer with its single adjacent layer.

We now apply this framework to a four-layer region of the sample. We first sweep the magnetic field along the $b$-axis from $B_b=-0.5$ T [Fig. \ref{fig:4L_Logic}a]. At $B_b=-0.5$ T, the $\mathrm{DR/R}$ spectrum [Fig. \ref{fig:4L_Logic}b] contains the surface and bulk exciton resonances associated with the fully FM configuration, indicating that all four layers are aligned parallel. At $B_b=-0.37$ T, an abrupt change in the spectrum occurs. The $X_{b,FM}$ resonance disappears and two bulk-related resonances emerge: $X_{b,AFM}$ and an additional resonance, $X_{inter}$, located approximately 8 meV below $X_{b,AFM}$. The absence of $X_{b,FM}$ indicates that the magnetization reversal involves an inner layer rather than a surface layer, as illustrated in Fig. \ref{fig:4L_Logic}a. Within our model, the resulting configuration contains an inner layer whose two neighboring layers have opposite magnetization directions, giving rise to the $X_{inter}$ resonance. The microscopic origin of $X_{inter}$ remains to be established.\\
The corresponding surface exciton response provides an additional signature of this configuration. Although both $X_{S,FM}$ and $X_{S,AFM}$ are expected, the latter is not readily apparent in the measured spectrum because of its substantially lower oscillator strength. To account for this response, we use a transfer-matrix model in which the permittivity of each layer is described by Lorentzian oscillators whose parameters depend on the relative magnetization of adjacent layers (see Methods). The model reproduces the measured spectrum and confirms the presence of a weak $X_{S,AFM}$ resonance beneath the dominant $X_{S,FM}$ feature. Fig. S6 displays exemplary transfer-matrix fits for the spectra discussed here. We consistently observe a reduced oscillator strength for one of the two surface-layer excitons across the sample, which we attribute to differences in interfacial quality. In particular, substrate roughness and interfacial contamination are known to affect the optical and electronic properties of two-dimensional materials \cite{raja2019dielectric,dean2010boron,cadiz2017excitonic} and may account for the reduced visibility.\\
Upon increasing the field from $B_b=-0.335$ T to $-0.325$ T, the spectrum evolves continuously [Fig. \ref{fig:4L_Logic}d] until it reaches the fully AFM configuration. Based on the spectral fingerprints established above, we attribute this gradual evolution to a domain-wall-mediated reversal of one of the surface-layer magnetizations, completing the transition from the intermediate configuration to the fully AFM state. 

We next apply the same approach to a five-layer region to test whether the spectroscopic fingerprints established above can be used to reconstruct more complex multilayer magnetic configurations. The magnetic-field-dependent $\mathrm{DR/R}$ response in Fig. \ref{fig:5L_Logic}a exhibits four abrupt changes, corresponding to successive changes in the underlying magnetization. Fig. \ref{fig:5L_Logic}b summarizes the magnetic configurations consistent with the observed excitonic resonances.


\begin{figure*}[t!]
\includegraphics[width = 17cm]{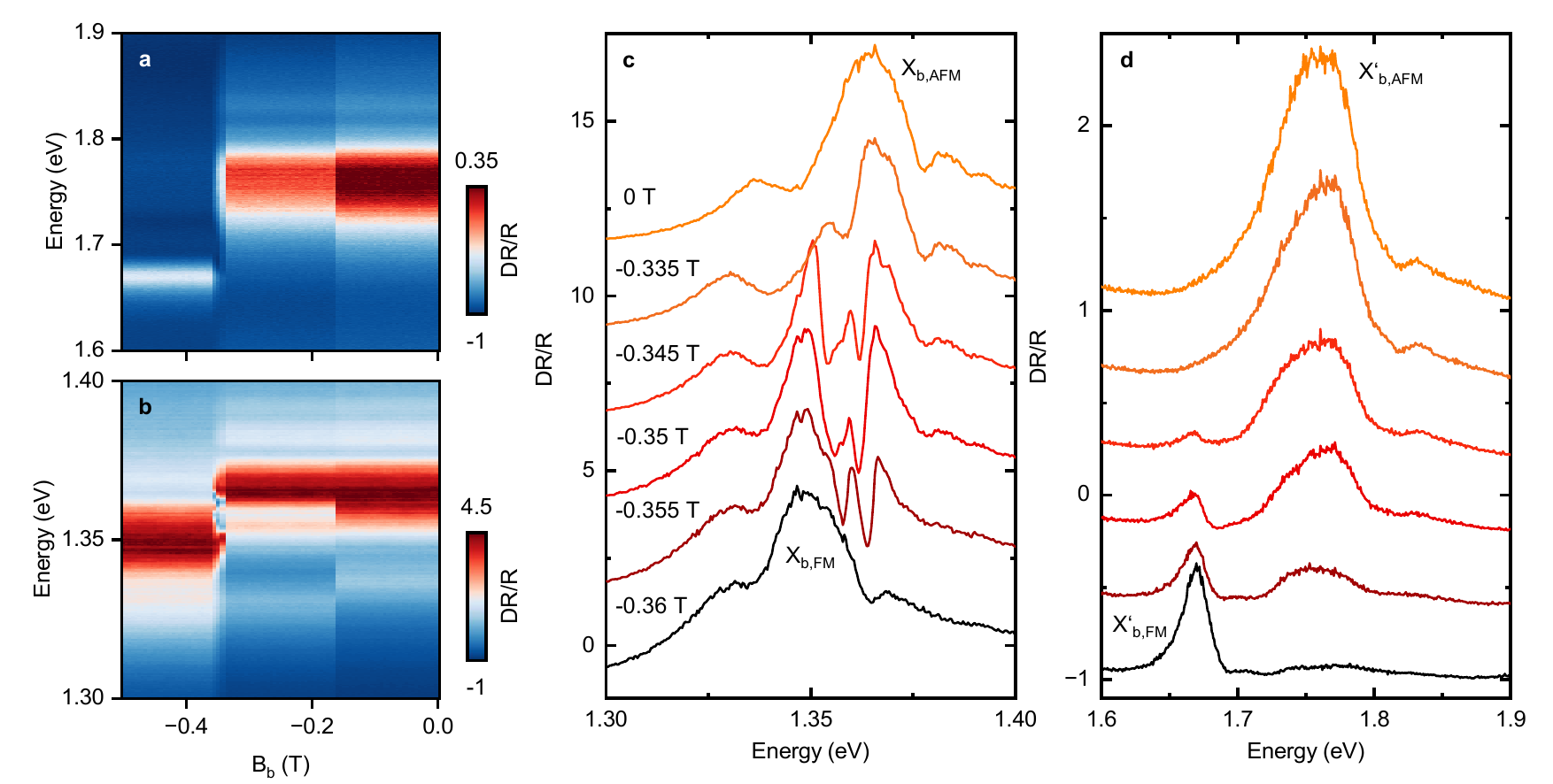}
\caption{\textbf{Intermediate resonances and higher energy excitons.} \textbf{a,b} Magnetic field dependent DR/R measurement in the 9 layer region higher energy region (\textbf{a}) and the lower energy region (\textbf{b}). \textbf{c} DR/R spectra for selected magnetic fields of the sweep in the lower energy range. \textbf{d} DR/R spectra for the same magnetic field strengths as in \textbf{c} in the higher energy range.} 
\label{fig:9L_B-exciton}
\end{figure*}


Below the saturation field, the system is in the fully FM configuration. At $B_b=-0.36$ T, three bulk-related resonances are observed: $X_{b,FM}$, $X_{inter}$, and $X_{b,AFM}$, with $X_{inter}$ located 6–7 meV below $X_{b,AFM}$ [Fig. \ref{fig:5L_Logic}c]. As in the four-layer case, the simultaneous presence of these three resonances requires an intermediate layer whose neighboring layers have opposite magnetization directions. This constraint strongly limits the possible magnetic configurations and yields a unique configuration apart from the vertically symmetric solution.

At the next transition [Fig. \ref{fig:5L_Logic}d], $X_{b,FM}$ disappears while $X_{b,AFM}$ remains present and $X_{inter}$ shifts towards the energy observed for the four-layer configuration. We therefore assign this transition to a rearrangement of the two inner-layer magnetizations, resulting in a configuration containing two $X_{inter}$-hosting layers. The detailed microscopic mechanism of this transition is unclear, but might be connected to anti-phase domain-wall formation which has been observed in A-type antiferromagnets \cite{tschudin2024imaging, wang2025configurable}. The small difference in the energy of the two $X_{inter}$ resonances further suggests that their excitonic environments are not completely equivalent, an effect not included in our simplified nearest-neighbor model.

The subsequent transition at $B_b=-0.35$ T primarily redistributes the spectral weight between $X_{inter}$ and $X_{b,AFM}$ while leaving their energies largely unchanged [Fig. \ref{fig:5L_Logic}e]. We therefore associate this change with reversal of an outer layer. The appearance of both $X_{S,FM}$ and $X_{S,AFM}$ provides an additional signature of this configuration and demonstrates the sensitivity of the surface excitons to the layer-resolved magnetic state. Finally, at $B_b=-0.15$ T, the remaining surface-layer magnetization reverses, resulting in the fully AFM configuration, for which only excitonic resonances associated with antiparallel layer alignment remain.

These results demonstrate that the excitonic response provides a spectroscopic fingerprint of the underlying layer-resolved magnetic configuration, allowing successive magnetic states to be reconstructed even as the number of layers and possible configurations increases. The energies, linewidths and oscillator strengths obtained in our transfer-matrix analysis are summarized in Table S1 and S2 for the 4 and 5 layer region respectively.

The excitonic resonances around $1.36$ eV provide a clear spectroscopic fingerprint of the underlying magnetic configuration through the appearance and oscillator strength of resonances associated with FM, AFM, and intermediate layer configurations. We next compare this response with a higher-energy exciton that can be measured simultaneously by using an appropriate diffraction grating. Although both excitonic resonances respond to the same underlying magnetization changes—and therefore exhibit spectral changes at the same magnetic fields—their optical responses are markedly different.

The higher-energy exciton shares several characteristics with $X_b$. In particular, despite having a more pronounced Wannier character, it remains confined predominantly to a single layer in the AFM configuration and should therefore also have a surface counterpart. In a 9-layer region, the main resonance in this spectral range is accompanied by a weaker, blueshifted feature [Fig. \ref{fig:Domain_Flip_Sketch} b]. We find no evidence that this feature corresponds to a surface exciton. First, unlike the surface resonance in the lower-energy spectral range, it is blueshifted relative to the main bulk resonance. Second, its field-dependent energy shift is comparable to that of the bulk exciton, whereas the lower-energy surface exciton exhibits a shift of approximately half the bulk value. Fig. S7 shows magnetic field dependent DR/R measurements with field applied along the a-axis. We therefore cannot unambiguously identify a surface exciton associated with the higher-energy resonance. We suggest measurements at high magnetic fields as in \cite{choi2026bulk} to clarify the origin of this resonance in the context of a surface confined exciton.

We next ask whether the higher-energy exciton also resolves the intermediate magnetic configurations identified in the lower-energy spectral range. To this end, we perform magnetic-field-dependent $\mathrm{DR/R}$ measurements on a 9-layer region and compare the two spectral ranges directly [Fig. \ref{fig:9L_B-exciton}a,b]. Importantly, spectral changes in both energy ranges occur at the same magnetic fields, confirming that they originate from the same underlying magnetization switching events. However, the spectral fingerprints of these events are qualitatively different. Near the saturation field, the lower-energy range resolves distinct FM, AFM, and intermediate exciton resonances [Fig. \ref{fig:9L_B-exciton}c]. In fact, two intermediate resonances can be distinguished in the range $-0.355~\mathrm{T}\leq B_b\leq-0.345~\mathrm{T}$. Such multiple intermediate resonances have previously been attributed to interference effects in substantially thicker CrSBr samples \cite{lopion2025optical}. Given the much smaller thickness of our 9-layer sample, we instead attribute their energy difference to intermediate excitons experiencing different dielectric environments and degrees of exciton localization, corresponding to different numbers of neighboring FM-aligned layers.

This observation also highlights a limitation of reconstructing the complete layer-resolved magnetization from the optical response. The simplified model used above assumes a unique exciton energy for a given relative alignment of neighboring layers, whereas in reality several intermediate excitonic resonances may coexist at slightly different energies. As the number of layers increases, the number of possible magnetic configurations and associated dielectric environments grows accordingly. A complete reconstruction of the magnetization from these closely spaced intermediate resonances therefore requires a more detailed microscopic model and is beyond the scope of this work.

In striking contrast, the higher-energy spectral range does not exhibit clearly resolved intermediate resonances [Fig. \ref{fig:9L_B-exciton}d]. Instead, the magnetic-field-driven changes appear predominantly as a stepwise transfer of oscillator strength between the FM- and AFM-associated resonances. Although the spectral line shape changes slightly during the transitions, we find no unambiguous evidence for additional intermediate excitonic states. In the fully FM configuration, the response further evolves into a group of closely spaced resonances, consistent with observations reported previously \cite{smiertka2026distinct}. Thus, while both excitonic manifolds respond to the same magnetic switching events, they provide qualitatively different optical fingerprints of the underlying magnetic configuration.

\section{Conclusion}
\indent \textit{In conclusion}, we have investigated how magnetization dynamics and layer-dependent magnetic configurations are encoded in the optical response of 4–9-layer CrSBr using low-temperature differential reflectance contrast ($\mathrm{DR/R}$) spectroscopy. We show that magnetization changes occurring through different mechanisms produce distinct spectroscopic signatures. Domain-wall motion through the optical spot leads to continuous spectral evolution that can be understood as a spatially weighted superposition of the responses of the two adjacent magnetic domains, whereas abrupt large-area magnetization reversal produces a correspondingly abrupt change in the optical response. Magnetic-field-dependent linescans across different regions of the sample further reveal substantial variations in domain-wall dynamics and provide a direct connection between spatial magnetic configurations and their optical fingerprints.

By combining the responses of bulk and surface excitons, we identify the layer-dependent magnetic configurations of 4- and 5-layer CrSBr consistent with the spectra. In particular, the appearance and energy of intermediate exciton resonances provide evidence for layers whose adjacent magnetizations are oppositely aligned, enabling successive magnetic configurations to be identified during the transition from ferromagnetic to antiferromagnetic order. At the same time, the observation of multiple intermediate resonances highlights the increasing complexity of the optical response with layer number and the limitations of assigning a unique exciton energy to each magnetic configuration.

A comparison with a higher-energy exciton further demonstrates that the sensitivity to magnetic order is strongly exciton-dependent. While the low-energy excitonic manifold exhibits distinct FM, AFM, intermediate, and surface-related resonances, the higher-energy exciton predominantly responds through a transfer of oscillator strength between FM- and AFM-associated features without clearly resolving intermediate configurations. These results establish excitonic spectroscopy as a sensitive optical probe of magnetic configurations and switching pathways in layered CrSBr, while highlighting the importance of exciton character and localization in determining the information encoded in the optical response.

\section{Methods}

\subsection{Sample fabrication:}
\indent Bulk CrSBr crystals were fabricated through chemical vapor transport \cite{klein2022control}. Nanometer thin CrSBr flakes were mechanically exfoliated onto Si substrates with an 80 nm thick oxide layer. The layer thicknesses were determined using an atomic force microscope (Oxford Instruments Cypher) equipped with AC160 cantilevers (Oxford Instruments). Atomic force microscopy (see scans in Fig. S1) yields CrSBr thicknesses ranging from 4 layers to 9 layers $h_{4L} = 3.3 \pm 0.2 \: \text{nm}$, $h_{5L} = 4.0 \pm 0.2 \: \text{nm}$, $h_{6L} = 4.8 \pm 0.2 \: \text{nm}$, $h_{7L} = 5.7 \pm 0.2 \: \text{nm}$, $h_{8L} = 6.5 \pm 0.2 \: \text{nm}$ and $h_{9L} = 7.3 \pm 0.2 \: \text{nm}$ .

\subsection{Optical spectroscopy:}
\indent Optical spectroscopy was carried out in a home-built, fiber based confocal setup for magneto-optical spectroscopy \cite{shree2021guide}. The sample was placed inside a closed-cycle cryostat (attocube systems, AttoDry 1000XL) equipped with a vector magnet (z-axis: solenoid, maximum field 5 T, x-/y-axis: Split Coil, maximum field 2 T). We used low temperature piezo-positioners (attocube systems, ANPx101 and ANPz102) to position the sample with respect to a low temperature apochromatic objective. DR/R measurements were performed in backscattering geometry at a sample temperature of 4.7 K. The signal was dispersed inside a Czerny-Turner spectrograph (Teledyne Princeton Instruments, SpectraPro HRS-500) and detected by a CCD-camera (Teledyne Princeton Instruments, Pylon BRexcelon 100). For DR/R measurements we used a Tungsten-Halogen lamp (Thorlabs, SLS201L/M) polarized along the crystal b-axis by a nanoparticle-film polarizer and an achromatic half-waveplate. Magnetic field dependent measurements were performed initializing the CrSBr sample in the FM-state, ramping the magnet to -0.5 T sample, followed by a sweep to 0.5 T in steps of 5-10 mT. Linescans were recorded in the same way, using single step movement of the piezos at 18 V applied, resulting in $\approx$ 50 nm wide steps.

\subsection{Transfer-matrix analysis}
\indent For the analysis of differential reflectance contrast measurements, we applied a transfer-matrix formalism \cite{dirnberger2023magneto, wang2023magnetically, robert2018optical}, using a Lorentzian oscillator model for the dielectric constant of CrSBr 
\begin{equation}
    \epsilon(\omega) = \epsilon_{\infty} + \sum_{j} \frac{f_{j}/\hbar^{2}}{\omega_{j}^{2} - \omega^{2} - i\Gamma_{j} \omega}
\end{equation}
where $\omega_{j}$ and $\Gamma_{j}$ denote the oscillator frequency and decay rate and $f_{j}$ denotes the oscillator strength of the $jth$ oscillator. We account for a constant background permittivity $\epsilon_{\infty} = 11.1$, similar to Wang \textit{et al.} \cite{wang2023magnetically} and used the permittivities $\epsilon_{SiO_{2}}$ from \cite{malitson1965interspecimen} and $\epsilon_{Si}$ from \cite{schinke2015uncertainty}. As described in the main text, for each specific three-layer-magnetization configuration, we used a separate set of oscillator parameters.

\section{Associated Content}
\subsection{Supporting Information}

\vspace{8pt}
\indent \textbf{Acknowledgements:} \\
Z.S. was supported by ERC-CZ program (project LL2101) from Ministry of Education Youth and Sports (MEYS) and by the project Advanced Functional Nanorobots (reg. No. CZ.02.1.01/0.0/0.0/15$_{-}$003/0000444 financed by the EFRR).\\

\indent \textbf{Author Contributions:} 
K.M. and Z.S. grew bulk CrSBr crystals. R.T. fabricated the CrSBr sample. L.K. and J.C. performed optical spectroscopy. P.M., L.K., R.T., R.v.K. performed and analysed AFM measurements. L.K., J.C. and B.U. analyzed the optical spectra. L.K., J.C. and B.U. discussed the results. B.U. suggested the experiments and supervised the project. L.K., and B.U. wrote the manuscript. All authors contributed to the final manuscript.\\

\indent \textbf{Competing interests}: The authors declare no competing interests.\\

\vspace{8pt}


%

\setcounter{figure}{0}
\renewcommand{\thefigure}{S\arabic{figure}}

\begin{figure*}[t]
\centering
\includegraphics[scale=1]{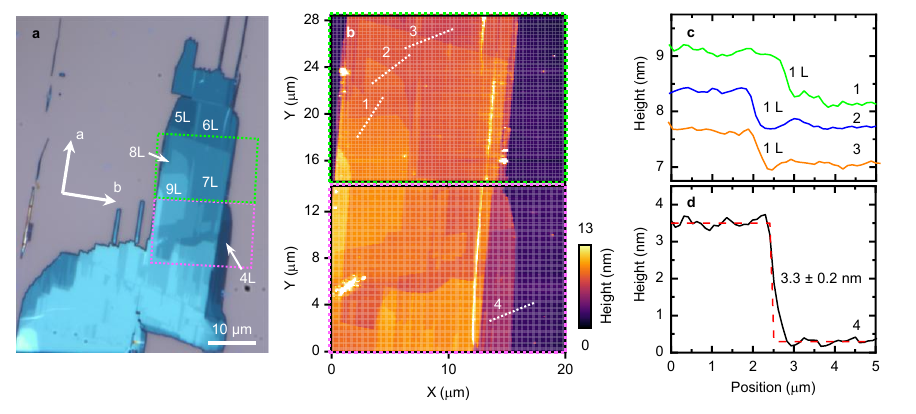}
\caption{\textbf{Sample characterization.} \textbf{a} Microscope image of the sample. \textbf{b} Atomic force microscope scan of the sample region indicated in \textbf{a}. \textbf{c,d} Atomic force microscope height profiles indicated by white dotted lines in \textbf{b}.} 
\label{fig:Sup1}
\end{figure*}

\begin{figure*}[t]
\includegraphics[scale=1]{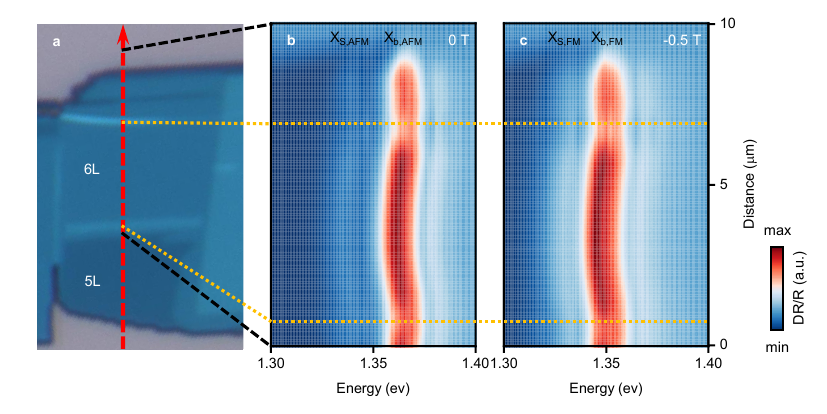}
\caption{\textbf{Linecuts in the fully AFM and FM state.} \textbf{a} Microscope image of the sample region across which linescan from maintext indicated by the red arrow were performed. Dashed yellow lines indicate sample features. \textbf{b} DR/R linescan at $B_b = 0$ T in the fully AFM state and \textbf{b} at $B_b = -0.5$ T in the fully FM state.} 
\label{fig:Sup2}
\end{figure*}

\begin{figure*}[t]
\includegraphics[scale=1]{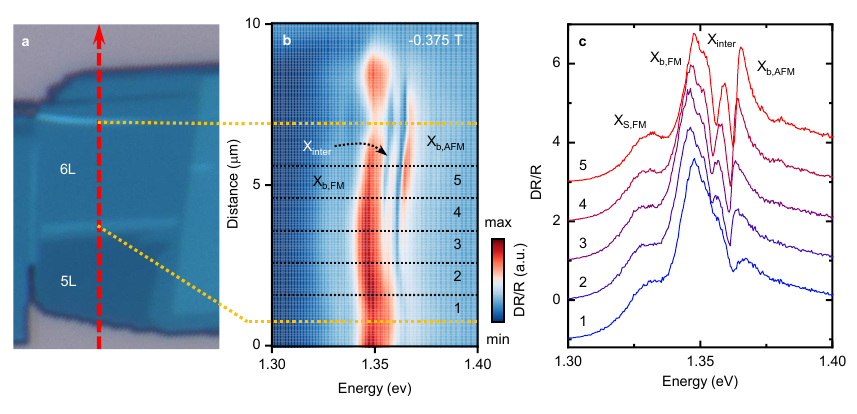}
\caption{\textbf{Spectral signatures of the domain wall in the 6 layers region.} \textbf{a} Microscope image of the sample region across which the linescans in the maintext were performed. Red arrow indicates the linescan path. \textbf{b} DR/R linescan from the maintext at $B_b = -0.375$ T. Dashed yellow lines indicate sample features. \textbf{c} DR/R spectra corresponding to linecuts marked by black dashed lines in \textbf{b}.} 
\label{fig:Sup3}
\end{figure*}

\begin{figure*}[t]
\includegraphics[scale=1]{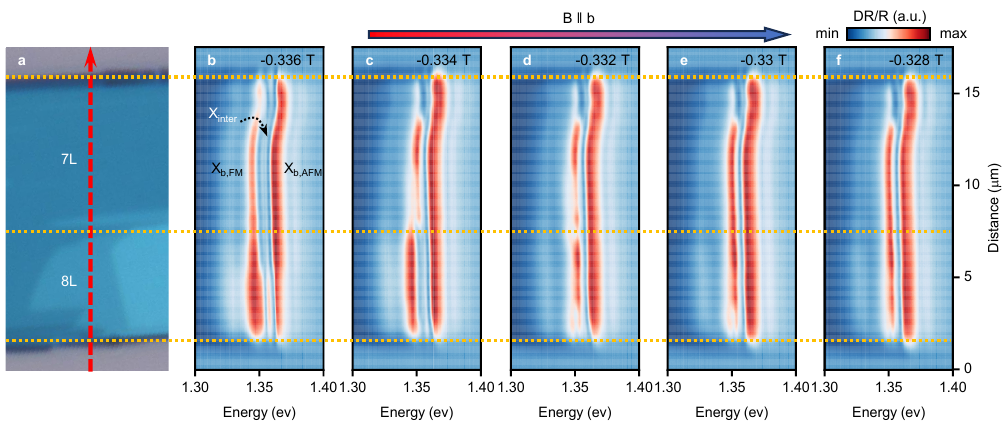}
\caption{\textbf{Domain wall extending along the b-axis.} \textbf{a} Microscope image of the sample region across which linescans indicated by the red arrow were performed. \textbf{b-f} DR/R linescans displaying a domain wall moving almost perpendicularly to the scan direction. Dashed yellow lines indicate sample borders and a sample specific region.} 
\label{fig:Sup4}
\end{figure*}

\begin{figure*}[t]
\includegraphics[scale=1]{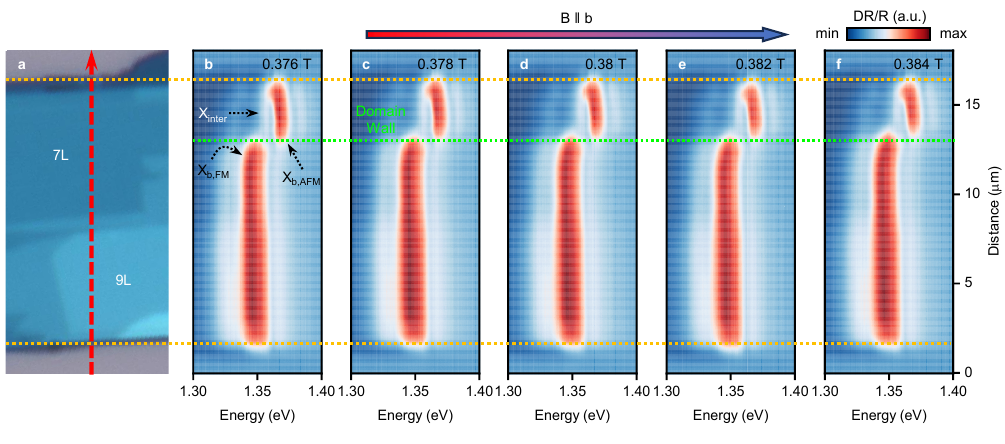}
\caption{\textbf{Domain wall pinning.} \textbf{a} Microscope image of the sample region across which linescans indicated by the red arrow were performed. \textbf{b-f} DR/R linescans displaying a domain wall dividing the 7 layers region. Region below domain wall is in fully FM state, region above the domain wall is almost fully in AFM state. Green dashed line indicates position of the domain wall for $B_b = 0.376$ T and serves as a guide for the eye. Dashed yellow lines indicate sample borders. The domain wall displays only minimal movement.} 
\label{fig:Sup5}
\end{figure*}

\begin{figure*}[t]
\includegraphics[scale=1]{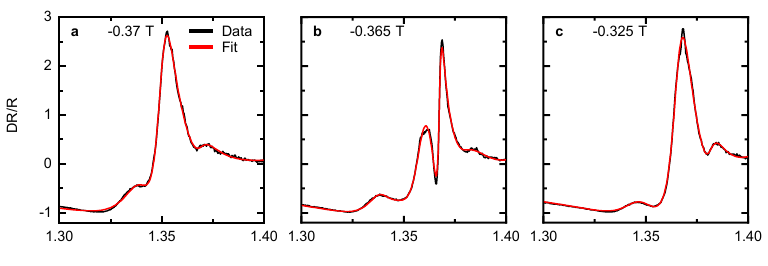}
\caption{\textbf{Transfer-matrix fits.} \textbf{a-c} Exemplary transfer-matrix fits for 4 layer data discussed in Fig. 3 of the main text.} 
\label{fig:Sup6}
\end{figure*}

\clearpage

\setcounter{table}{0}
\renewcommand{\thetable}{S\arabic{table}}

\begin{table*}[ht]
\centering
\begin{tabular}{l|ccc|ccc|ccc}
\hline
 &
 & $B_b = -0.37$ T &  &  & $B_b = -0.365$ T &  &  & $B_b = -0.325$ T &   \\
\hline
Resonance &
E (eV) & $\Gamma$ (meV) & f (eV$^2$) & E (eV) & $\Gamma$ (meV) & f (eV$^2$) & E (eV) & $\Gamma$ (meV) & f (eV$^2$) \\
\hline
$E_{b,AFM}$
& &  &  & 1.368 & 1 & 2.0 & 1.368 & 1.9 & 2.8  \\
$E_{S,AFM}$
&  &  &  & 1.347 & 14.3 & 0.7 & 1.346 & 17.8 & 1.1 \\
$E_{b,FM}$
& 1.352 & 1.9 & 3.0 &  &  &  &  &  &  \\
$E_{S,FM}$
& 1.339 & 14.8 & 1.3 & 1.338 & 12.6 & 1.9 &  &  &  \\
$E_{inter}$
&  &  &  & 1.3605 & 5.9 & 4.1 &  &  &  \\
\hline
\end{tabular}
\caption{Fit parameters obtained via transfer-matrix formalism for the DR/R measurement in Fig. 3 of the main text.}
\end{table*}

\begin{table*}[ht]
\centering
\begin{tabular}{l|ccc|ccc|ccc}
\hline
 &
 & $B_b = -0.37$ T &  &  & $B_b = -0.36$ T &  &  & $B_b = -0.35$ T &   \\
\hline
Resonance &
E (eV) & $\Gamma$ (meV) & f (eV$^2$) & E (eV) & $\Gamma$ (meV) & f (eV$^2$) & E (eV) & $\Gamma$ (meV) & f (eV$^2$) \\
\hline
$E_{b,AFM}$
& &  &  & 1.365 & 1.3 & 2.1 & 1.365 & 1.0 & 6.3  \\
$E_{S,AFM}$
&  &  &  & 1.341 & 14.1 & 1.7 &  &  &  \\
$E_{b,FM}$
& 1.348 & 1 & 3.7 & 1.348 & 1.2 & 5.0 &  &  &  \\
$E_{S,FM}$
& 1.335 & 10 & 1.5 & 1.333 & 8.7 & 2.0 & 1.334 & 11.2 & 1.6 \\
$E_{inter}$
&  &  &  & 1.358 & 3.1 & 2.7 & 1.356 & 5.1 & 1.8 \\
\hline
\end{tabular}
\begin{tabular}{l|ccc|ccc}
\hline
 &
 & $B_b = -0.15$ T &  &  & $B_b = 0$ T &   \\
\hline
Resonance &
E (eV) & $\Gamma$ (meV) & f (eV$^2$) & E (eV) & $\Gamma$ (meV) & f (eV$^2$) \\
\hline
$E_{b,AFM}$
& 1.364 & 1.0 & 4.1 & 1.364 & 1.0 & 3.4   \\
$E_{S,AFM}$
& 1.341 & 13.5 & 2.0 & 1.342 & 14.2 & 1.6 \\
$E_{b,FM}$
&  &  &  &  &  &    \\
$E_{S,FM}$
& 1.334 & 10.1 & 1.1 &  &  &  \\
$E_{inter}$
& 1.356 & 4.8 & 1.7 &  &  &   \\
\hline
\end{tabular}
\caption{Fit parameters obtained via transfer-matrix formalism for the DR/R measurement in Fig. 4 of the main text.}
\end{table*}

\begin{figure*}[t]
\includegraphics[scale=1]{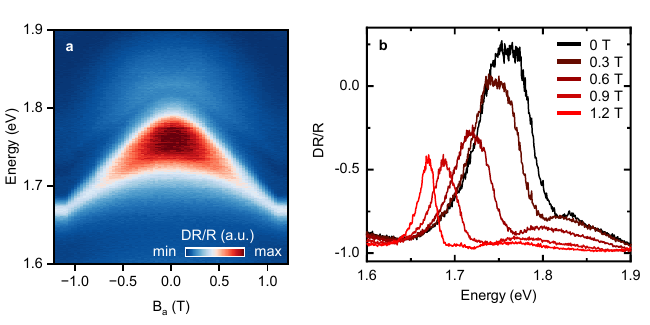}
\caption{\textbf{High energy excitons and spin canting.} \textbf{a} DR/R magnetic field sweep along the a-axis. \textbf{b} DR/R spectra of the measurement in \textbf{a} for selected magnetic field values.} 
\label{fig:Sup7}
\end{figure*}


\begin{thebibliography}{43}%
\makeatletter
\providecommand \@ifxundefined [1]{%
 \@ifx{#1\undefined}
}%
\providecommand \@ifnum [1]{%
 \ifnum #1\expandafter \@firstoftwo
 \else \expandafter \@secondoftwo
 \fi
}%
\providecommand \@ifx [1]{%
 \ifx #1\expandafter \@firstoftwo
 \else \expandafter \@secondoftwo
 \fi
}%
\providecommand \natexlab [1]{#1}%
\providecommand \enquote  [1]{``#1''}%
\providecommand \bibnamefont  [1]{#1}%
\providecommand \bibfnamefont [1]{#1}%
\providecommand \citenamefont [1]{#1}%
\providecommand \href@noop [0]{\@secondoftwo}%
\providecommand \href [0]{\begingroup \@sanitize@url \@href}%
\providecommand \@href[1]{\@@startlink{#1}\@@href}%
\providecommand \@@href[1]{\endgroup#1\@@endlink}%
\providecommand \@sanitize@url [0]{\catcode `\\12\catcode `\$12\catcode
  `\&12\catcode `\#12\catcode `\^12\catcode `\_12\catcode `\%12\relax}%
\providecommand \@@startlink[1]{}%
\providecommand \@@endlink[0]{}%
\providecommand \url  [0]{\begingroup\@sanitize@url \@url }%
\providecommand \@url [1]{\endgroup\@href {#1}{\urlprefix }}%
\providecommand \urlprefix  [0]{URL }%
\providecommand \Eprint [0]{\href }%
\providecommand \doibase [0]{https://doi.org/}%
\providecommand \selectlanguage [0]{\@gobble}%
\providecommand \bibinfo  [0]{\@secondoftwo}%
\providecommand \bibfield  [0]{\@secondoftwo}%
\providecommand \translation [1]{[#1]}%
\providecommand \BibitemOpen [0]{}%
\providecommand \bibitemStop [0]{}%
\providecommand \bibitemNoStop [0]{.\EOS\space}%
\providecommand \EOS [0]{\spacefactor3000\relax}%
\providecommand \BibitemShut  [1]{\csname bibitem#1\endcsname}%
\let\auto@bib@innerbib\@empty
\bibitem [{\citenamefont {Ahn}(2020)}]{ahn20202d}%
  \BibitemOpen
  \bibfield  {author} {\bibinfo {author} {\bibfnamefont {E.~C.}\ \bibnamefont
  {Ahn}},\ }\bibfield  {title} {\bibinfo {title} {2d materials for spintronic
  devices},\ } {\bibfield  {journal} {\bibinfo  {journal} {npj 2D
  Materials and Applications}\ }\textbf {\bibinfo {volume} {4}},\ \bibinfo
  {pages} {17} (\bibinfo {year} {2020})}\BibitemShut {NoStop}%
\bibitem [{\citenamefont {Mi}\ \emph {et~al.}(2023)\citenamefont {Mi},
  \citenamefont {Xiao}, \citenamefont {Yu}, \citenamefont {Zhang},
  \citenamefont {Wang}, \citenamefont {Cao},\ and\ \citenamefont
  {Wang}}]{mi2023two}%
  \BibitemOpen
  \bibfield  {author} {\bibinfo {author} {\bibfnamefont {M.}~\bibnamefont
  {Mi}}, \bibinfo {author} {\bibfnamefont {H.}~\bibnamefont {Xiao}}, \bibinfo
  {author} {\bibfnamefont {L.}~\bibnamefont {Yu}}, \bibinfo {author}
  {\bibfnamefont {Y.}~\bibnamefont {Zhang}}, \bibinfo {author} {\bibfnamefont
  {Y.}~\bibnamefont {Wang}}, \bibinfo {author} {\bibfnamefont {Q.}~\bibnamefont
  {Cao}},\ and\ \bibinfo {author} {\bibfnamefont {Y.}~\bibnamefont {Wang}},\
  }\bibfield  {title} {\bibinfo {title} {Two-dimensional magnetic materials for
  spintronic devices},\ } {\bibfield  {journal} {\bibinfo
  {journal} {Materials Today Nano}\ ,\ \bibinfo {pages} {100408}} (\bibinfo
  {year} {2023})}\BibitemShut {NoStop}%
\bibitem [{\citenamefont {Adak}\ \emph {et~al.}(2026)\citenamefont {Adak},
  \citenamefont {Dirnberger}, \citenamefont {Acharya}, \citenamefont {Kamra},
  \citenamefont {Xu},\ and\ \citenamefont {Menon}}]{adak2026excitons}%
  \BibitemOpen
  \bibfield  {author} {\bibinfo {author} {\bibfnamefont {P.~C.}\ \bibnamefont
  {Adak}}, \bibinfo {author} {\bibfnamefont {F.}~\bibnamefont {Dirnberger}},
  \bibinfo {author} {\bibfnamefont {S.}~\bibnamefont {Acharya}}, \bibinfo
  {author} {\bibfnamefont {A.}~\bibnamefont {Kamra}}, \bibinfo {author}
  {\bibfnamefont {X.}~\bibnamefont {Xu}},\ and\ \bibinfo {author}
  {\bibfnamefont {V.~M.}\ \bibnamefont {Menon}},\ }\bibfield  {title} {\bibinfo
  {title} {Excitons in van der waals magnetic materials},\ }
  {\bibfield  {journal} {\bibinfo  {journal} {Nature Materials}\ ,\ \bibinfo
  {pages} {1}} (\bibinfo {year} {2026})}\BibitemShut {NoStop}%
\bibitem [{\citenamefont {Park}\ \emph {et~al.}(2026)\citenamefont {Park},
  \citenamefont {Zhang}, \citenamefont {Cheong}, \citenamefont {Kim},
  \citenamefont {Belvin}, \citenamefont {Hsieh}, \citenamefont {Ning},\ and\
  \citenamefont {Gedik}}]{park20262d}%
  \BibitemOpen
  \bibfield  {author} {\bibinfo {author} {\bibfnamefont {J.-G.}\ \bibnamefont
  {Park}}, \bibinfo {author} {\bibfnamefont {K.-X.}\ \bibnamefont {Zhang}},
  \bibinfo {author} {\bibfnamefont {H.}~\bibnamefont {Cheong}}, \bibinfo
  {author} {\bibfnamefont {J.~H.}\ \bibnamefont {Kim}}, \bibinfo {author}
  {\bibfnamefont {C.~A.}\ \bibnamefont {Belvin}}, \bibinfo {author}
  {\bibfnamefont {D.}~\bibnamefont {Hsieh}}, \bibinfo {author} {\bibfnamefont
  {H.}~\bibnamefont {Ning}},\ and\ \bibinfo {author} {\bibfnamefont
  {N.}~\bibnamefont {Gedik}},\ }\bibfield  {title} {\bibinfo {title} {2d van
  der waals magnets: from fundamental physics to applications},\ }
  {\bibfield  {journal} {\bibinfo  {journal} {Reviews of Modern Physics}\
  }\textbf {\bibinfo {volume} {98}},\ \bibinfo {pages} {025003} (\bibinfo
  {year} {2026})}\BibitemShut {NoStop}%
\bibitem [{\citenamefont {Diederich}\ \emph {et~al.}(2023)\citenamefont
  {Diederich}, \citenamefont {Cenker}, \citenamefont {Ren}, \citenamefont
  {Fonseca}, \citenamefont {Chica}, \citenamefont {Bae}, \citenamefont {Zhu},
  \citenamefont {Roy}, \citenamefont {Cao}, \citenamefont {Xiao} \emph
  {et~al.}}]{diederich2023tunable}%
  \BibitemOpen
  \bibfield  {author} {\bibinfo {author} {\bibfnamefont {G.~M.}\ \bibnamefont
  {Diederich}}, \bibinfo {author} {\bibfnamefont {J.}~\bibnamefont {Cenker}},
  \bibinfo {author} {\bibfnamefont {Y.}~\bibnamefont {Ren}}, \bibinfo {author}
  {\bibfnamefont {J.}~\bibnamefont {Fonseca}}, \bibinfo {author} {\bibfnamefont
  {D.~G.}\ \bibnamefont {Chica}}, \bibinfo {author} {\bibfnamefont {Y.~J.}\
  \bibnamefont {Bae}}, \bibinfo {author} {\bibfnamefont {X.}~\bibnamefont
  {Zhu}}, \bibinfo {author} {\bibfnamefont {X.}~\bibnamefont {Roy}}, \bibinfo
  {author} {\bibfnamefont {T.}~\bibnamefont {Cao}}, \bibinfo {author}
  {\bibfnamefont {D.}~\bibnamefont {Xiao}}, \emph {et~al.},\ }\bibfield
  {title} {\bibinfo {title} {Tunable interaction between excitons and
  hybridized magnons in a layered semiconductor},\ } {\bibfield
  {journal} {\bibinfo  {journal} {Nature Nanotechnology}\ }\textbf {\bibinfo
  {volume} {18}},\ \bibinfo {pages} {23} (\bibinfo {year} {2023})}\BibitemShut
  {NoStop}%
\bibitem [{\citenamefont {Datta}\ \emph {et~al.}(2025)\citenamefont {Datta},
  \citenamefont {Adak}, \citenamefont {Yu}, \citenamefont {Dharmapalan},
  \citenamefont {Hall}, \citenamefont {Vakulenko}, \citenamefont
  {Komissarenko}, \citenamefont {Kurganov}, \citenamefont {Quan}, \citenamefont
  {Wang} \emph {et~al.}}]{datta2025magnon}%
  \BibitemOpen
  \bibfield  {author} {\bibinfo {author} {\bibfnamefont {B.}~\bibnamefont
  {Datta}}, \bibinfo {author} {\bibfnamefont {P.~C.}\ \bibnamefont {Adak}},
  \bibinfo {author} {\bibfnamefont {S.}~\bibnamefont {Yu}}, \bibinfo {author}
  {\bibfnamefont {A.~V.}\ \bibnamefont {Dharmapalan}}, \bibinfo {author}
  {\bibfnamefont {S.~J.}\ \bibnamefont {Hall}}, \bibinfo {author}
  {\bibfnamefont {A.}~\bibnamefont {Vakulenko}}, \bibinfo {author}
  {\bibfnamefont {F.}~\bibnamefont {Komissarenko}}, \bibinfo {author}
  {\bibfnamefont {E.}~\bibnamefont {Kurganov}}, \bibinfo {author}
  {\bibfnamefont {J.}~\bibnamefont {Quan}}, \bibinfo {author} {\bibfnamefont
  {W.}~\bibnamefont {Wang}}, \emph {et~al.},\ }\bibfield  {title} {\bibinfo
  {title} {Magnon-mediated exciton–exciton interaction in a van der waals
  antiferromagnet},\ } {\bibfield  {journal} {\bibinfo  {journal}
  {Nat. Mater.}\ } (\bibinfo {year} {2025})}\BibitemShut {NoStop}%
\bibitem [{\citenamefont {Bae}\ \emph {et~al.}(2022)\citenamefont {Bae},
  \citenamefont {Wang}, \citenamefont {Scheie}, \citenamefont {Xu},
  \citenamefont {Chica}, \citenamefont {Diederich}, \citenamefont {Cenker},
  \citenamefont {Ziebel}, \citenamefont {Bai}, \citenamefont {Ren} \emph
  {et~al.}}]{bae2022exciton}%
  \BibitemOpen
  \bibfield  {author} {\bibinfo {author} {\bibfnamefont {Y.~J.}\ \bibnamefont
  {Bae}}, \bibinfo {author} {\bibfnamefont {J.}~\bibnamefont {Wang}}, \bibinfo
  {author} {\bibfnamefont {A.}~\bibnamefont {Scheie}}, \bibinfo {author}
  {\bibfnamefont {J.}~\bibnamefont {Xu}}, \bibinfo {author} {\bibfnamefont
  {D.~G.}\ \bibnamefont {Chica}}, \bibinfo {author} {\bibfnamefont {G.~M.}\
  \bibnamefont {Diederich}}, \bibinfo {author} {\bibfnamefont {J.}~\bibnamefont
  {Cenker}}, \bibinfo {author} {\bibfnamefont {M.~E.}\ \bibnamefont {Ziebel}},
  \bibinfo {author} {\bibfnamefont {Y.}~\bibnamefont {Bai}}, \bibinfo {author}
  {\bibfnamefont {H.}~\bibnamefont {Ren}}, \emph {et~al.},\ }\bibfield  {title}
  {\bibinfo {title} {Exciton-coupled coherent magnons in a 2d semiconductor},\
  } {\bibfield  {journal} {\bibinfo  {journal} {Nature}\ }\textbf
  {\bibinfo {volume} {609}},\ \bibinfo {pages} {282} (\bibinfo {year}
  {2022})}\BibitemShut {NoStop}%
\bibitem [{\citenamefont {Lin}\ \emph {et~al.}(2024)\citenamefont {Lin},
  \citenamefont {Sun}, \citenamefont {Dirnberger}, \citenamefont {Li},
  \citenamefont {Qu}, \citenamefont {Wen}, \citenamefont {Sofer}, \citenamefont
  {S\"oll}, \citenamefont {Winnerl}, \citenamefont {Helm} \emph
  {et~al.}}]{lin2024strong}%
  \BibitemOpen
  \bibfield  {author} {\bibinfo {author} {\bibfnamefont {K.}~\bibnamefont
  {Lin}}, \bibinfo {author} {\bibfnamefont {X.}~\bibnamefont {Sun}}, \bibinfo
  {author} {\bibfnamefont {F.}~\bibnamefont {Dirnberger}}, \bibinfo {author}
  {\bibfnamefont {Y.}~\bibnamefont {Li}}, \bibinfo {author} {\bibfnamefont
  {J.}~\bibnamefont {Qu}}, \bibinfo {author} {\bibfnamefont {P.}~\bibnamefont
  {Wen}}, \bibinfo {author} {\bibfnamefont {Z.}~\bibnamefont {Sofer}}, \bibinfo
  {author} {\bibfnamefont {A.}~\bibnamefont {S\"oll}}, \bibinfo {author}
  {\bibfnamefont {S.}~\bibnamefont {Winnerl}}, \bibinfo {author} {\bibfnamefont
  {M.}~\bibnamefont {Helm}}, \emph {et~al.},\ }\bibfield  {title} {\bibinfo
  {title} {Strong exciton--phonon coupling as a fingerprint of magnetic
  ordering in van der waals layered {CrSBr}},\ } {\bibfield
  {journal} {\bibinfo  {journal} {ACS nano}\ }\textbf {\bibinfo {volume}
  {18}},\ \bibinfo {pages} {2898} (\bibinfo {year} {2024})}\BibitemShut
  {NoStop}%
\bibitem [{\citenamefont {Dirnberger}\ \emph {et~al.}(2023)\citenamefont
  {Dirnberger}, \citenamefont {Quan}, \citenamefont {Bushati}, \citenamefont
  {Diederich}, \citenamefont {Florian}, \citenamefont {Klein}, \citenamefont
  {Mosina}, \citenamefont {Sofer}, \citenamefont {Xu}, \citenamefont {Kamra}
  \emph {et~al.}}]{dirnberger2023magneto}%
  \BibitemOpen
  \bibfield  {author} {\bibinfo {author} {\bibfnamefont {F.}~\bibnamefont
  {Dirnberger}}, \bibinfo {author} {\bibfnamefont {J.}~\bibnamefont {Quan}},
  \bibinfo {author} {\bibfnamefont {R.}~\bibnamefont {Bushati}}, \bibinfo
  {author} {\bibfnamefont {G.~M.}\ \bibnamefont {Diederich}}, \bibinfo {author}
  {\bibfnamefont {M.}~\bibnamefont {Florian}}, \bibinfo {author} {\bibfnamefont
  {J.}~\bibnamefont {Klein}}, \bibinfo {author} {\bibfnamefont
  {K.}~\bibnamefont {Mosina}}, \bibinfo {author} {\bibfnamefont
  {Z.}~\bibnamefont {Sofer}}, \bibinfo {author} {\bibfnamefont
  {X.}~\bibnamefont {Xu}}, \bibinfo {author} {\bibfnamefont {A.}~\bibnamefont
  {Kamra}}, \emph {et~al.},\ }\bibfield  {title} {\bibinfo {title}
  {Magneto-optics in a van der waals magnet tuned by self-hybridized
  polaritons},\ } {\bibfield  {journal} {\bibinfo  {journal}
  {Nature}\ }\textbf {\bibinfo {volume} {620}},\ \bibinfo {pages} {533}
  (\bibinfo {year} {2023})}\BibitemShut {NoStop}%
\bibitem [{\citenamefont {Wang}\ \emph {et~al.}(2023)\citenamefont {Wang},
  \citenamefont {Zhang}, \citenamefont {Yang}, \citenamefont {Lin},
  \citenamefont {Chen}, \citenamefont {Yang}, \citenamefont {Gong},
  \citenamefont {Chen}, \citenamefont {Ye},\ and\ \citenamefont
  {Liu}}]{wang2023magnetically}%
  \BibitemOpen
  \bibfield  {author} {\bibinfo {author} {\bibfnamefont {T.}~\bibnamefont
  {Wang}}, \bibinfo {author} {\bibfnamefont {D.}~\bibnamefont {Zhang}},
  \bibinfo {author} {\bibfnamefont {S.}~\bibnamefont {Yang}}, \bibinfo {author}
  {\bibfnamefont {Z.}~\bibnamefont {Lin}}, \bibinfo {author} {\bibfnamefont
  {Q.}~\bibnamefont {Chen}}, \bibinfo {author} {\bibfnamefont {J.}~\bibnamefont
  {Yang}}, \bibinfo {author} {\bibfnamefont {Q.}~\bibnamefont {Gong}}, \bibinfo
  {author} {\bibfnamefont {Z.}~\bibnamefont {Chen}}, \bibinfo {author}
  {\bibfnamefont {Y.}~\bibnamefont {Ye}},\ and\ \bibinfo {author}
  {\bibfnamefont {W.}~\bibnamefont {Liu}},\ }\bibfield  {title} {\bibinfo
  {title} {Magnetically-dressed {CrSBr} exciton-polaritons in ultrastrong
  coupling regime},\ } {\bibfield  {journal} {\bibinfo  {journal}
  {Nature Communications}\ }\textbf {\bibinfo {volume} {14}},\ \bibinfo {pages}
  {5966} (\bibinfo {year} {2023})}\BibitemShut {NoStop}%
\bibitem [{\citenamefont {Budak}\ \emph {et~al.}(2026)\citenamefont {Budak},
  \citenamefont {Riedel}, \citenamefont {Kamra}, \citenamefont {Rinke},
  \citenamefont {Back}, \citenamefont {Stosiek},\ and\ \citenamefont
  {Dirnberger}}]{budak2026role}%
  \BibitemOpen
  \bibfield  {author} {\bibinfo {author} {\bibfnamefont {G.}~\bibnamefont
  {Budak}}, \bibinfo {author} {\bibfnamefont {C.}~\bibnamefont {Riedel}},
  \bibinfo {author} {\bibfnamefont {A.}~\bibnamefont {Kamra}}, \bibinfo
  {author} {\bibfnamefont {P.}~\bibnamefont {Rinke}}, \bibinfo {author}
  {\bibfnamefont {C.}~\bibnamefont {Back}}, \bibinfo {author} {\bibfnamefont
  {M.}~\bibnamefont {Stosiek}},\ and\ \bibinfo {author} {\bibfnamefont
  {F.}~\bibnamefont {Dirnberger}},\ }\bibfield  {title} {\bibinfo {title} {Role
  of photonic interference in exciton-mediated magneto-optic responses},\
  } {\bibfield  {journal} {\bibinfo  {journal} {Physical Review
  B}\ }\textbf {\bibinfo {volume} {114}},\ \bibinfo {pages} {094405} (\bibinfo
  {year} {2026})}\BibitemShut {NoStop}%
\bibitem [{\citenamefont {Li}\ \emph {et~al.}(2026)\citenamefont {Li},
  \citenamefont {Sun}, \citenamefont {Wang}, \citenamefont {Wang},
  \citenamefont {Watanabe}, \citenamefont {Taniguchi}, \citenamefont {Liu},
  \citenamefont {Yu}, \citenamefont {Li}, \citenamefont {Zhang} \emph
  {et~al.}}]{li2026electric}%
  \BibitemOpen
  \bibfield  {author} {\bibinfo {author} {\bibfnamefont {X.}~\bibnamefont
  {Li}}, \bibinfo {author} {\bibfnamefont {Y.}~\bibnamefont {Sun}}, \bibinfo
  {author} {\bibfnamefont {X.}~\bibnamefont {Wang}}, \bibinfo {author}
  {\bibfnamefont {C.}~\bibnamefont {Wang}}, \bibinfo {author} {\bibfnamefont
  {K.}~\bibnamefont {Watanabe}}, \bibinfo {author} {\bibfnamefont
  {T.}~\bibnamefont {Taniguchi}}, \bibinfo {author} {\bibfnamefont
  {S.}~\bibnamefont {Liu}}, \bibinfo {author} {\bibfnamefont {T.}~\bibnamefont
  {Yu}}, \bibinfo {author} {\bibfnamefont {L.}~\bibnamefont {Li}}, \bibinfo
  {author} {\bibfnamefont {T.}~\bibnamefont {Zhang}}, \emph {et~al.},\
  }\bibfield  {title} {\bibinfo {title} {Electric-field-driven magnetic
  switching and tightly bound interlayer excitons in bilayer {CrSBr}},\
  } {\bibfield  {journal} {\bibinfo  {journal} {arXiv preprint
  arXiv:2606.20159}\ } (\bibinfo {year} {2026})}\BibitemShut {NoStop}%
\bibitem [{\citenamefont {Wilson}\ \emph {et~al.}(2021)\citenamefont {Wilson},
  \citenamefont {Lee}, \citenamefont {Cenker}, \citenamefont {Xie},
  \citenamefont {Dismukes}, \citenamefont {Telford}, \citenamefont {Fonseca},
  \citenamefont {Sivakumar}, \citenamefont {Dean}, \citenamefont {Cao} \emph
  {et~al.}}]{wilson2021interlayer}%
  \BibitemOpen
  \bibfield  {author} {\bibinfo {author} {\bibfnamefont {N.~P.}\ \bibnamefont
  {Wilson}}, \bibinfo {author} {\bibfnamefont {K.}~\bibnamefont {Lee}},
  \bibinfo {author} {\bibfnamefont {J.}~\bibnamefont {Cenker}}, \bibinfo
  {author} {\bibfnamefont {K.}~\bibnamefont {Xie}}, \bibinfo {author}
  {\bibfnamefont {A.~H.}\ \bibnamefont {Dismukes}}, \bibinfo {author}
  {\bibfnamefont {E.~J.}\ \bibnamefont {Telford}}, \bibinfo {author}
  {\bibfnamefont {J.}~\bibnamefont {Fonseca}}, \bibinfo {author} {\bibfnamefont
  {S.}~\bibnamefont {Sivakumar}}, \bibinfo {author} {\bibfnamefont
  {C.}~\bibnamefont {Dean}}, \bibinfo {author} {\bibfnamefont {T.}~\bibnamefont
  {Cao}}, \emph {et~al.},\ }\bibfield  {title} {\bibinfo {title} {Interlayer
  electronic coupling on demand in a 2d magnetic semiconductor},\ }\href@noop
  {} {\bibfield  {journal} {\bibinfo  {journal} {Nature Materials}\ }\textbf
  {\bibinfo {volume} {20}},\ \bibinfo {pages} {1657} (\bibinfo {year}
  {2021})}\BibitemShut {NoStop}%
\bibitem [{\citenamefont {Krelle}\ \emph {et~al.}(2025)\citenamefont {Krelle},
  \citenamefont {Tan}, \citenamefont {Markina}, \citenamefont {Mondal},
  \citenamefont {Mosina}, \citenamefont {Hagmann}, \citenamefont {von
  Klitzing}, \citenamefont {Watanabe}, \citenamefont {Taniguchi}, \citenamefont
  {Sofer} \emph {et~al.}}]{krelle2025magnetic}%
  \BibitemOpen
  \bibfield  {author} {\bibinfo {author} {\bibfnamefont {L.}~\bibnamefont
  {Krelle}}, \bibinfo {author} {\bibfnamefont {R.}~\bibnamefont {Tan}},
  \bibinfo {author} {\bibfnamefont {D.}~\bibnamefont {Markina}}, \bibinfo
  {author} {\bibfnamefont {P.}~\bibnamefont {Mondal}}, \bibinfo {author}
  {\bibfnamefont {K.}~\bibnamefont {Mosina}}, \bibinfo {author} {\bibfnamefont
  {K.}~\bibnamefont {Hagmann}}, \bibinfo {author} {\bibfnamefont
  {R.}~\bibnamefont {von Klitzing}}, \bibinfo {author} {\bibfnamefont
  {K.}~\bibnamefont {Watanabe}}, \bibinfo {author} {\bibfnamefont
  {T.}~\bibnamefont {Taniguchi}}, \bibinfo {author} {\bibfnamefont
  {Z.}~\bibnamefont {Sofer}}, \emph {et~al.},\ }\bibfield  {title} {\bibinfo
  {title} {Magnetic correlation spectroscopy in {CrSBr}},\ }
  {\bibfield  {journal} {\bibinfo  {journal} {ACS nano}\ }\textbf {\bibinfo
  {volume} {19}},\ \bibinfo {pages} {33156} (\bibinfo {year}
  {2025})}\BibitemShut {NoStop}%
\bibitem [{\citenamefont {Tabataba-Vakili}\ \emph {et~al.}(2024)\citenamefont
  {Tabataba-Vakili}, \citenamefont {Nguyen}, \citenamefont {Rupp},
  \citenamefont {Mosina}, \citenamefont {Papavasileiou}, \citenamefont
  {Watanabe}, \citenamefont {Taniguchi}, \citenamefont {Maletinsky},
  \citenamefont {Glazov}, \citenamefont {Sofer} \emph
  {et~al.}}]{tabataba2024doping}%
  \BibitemOpen
  \bibfield  {author} {\bibinfo {author} {\bibfnamefont {F.}~\bibnamefont
  {Tabataba-Vakili}}, \bibinfo {author} {\bibfnamefont {H.~P.}\ \bibnamefont
  {Nguyen}}, \bibinfo {author} {\bibfnamefont {A.}~\bibnamefont {Rupp}},
  \bibinfo {author} {\bibfnamefont {K.}~\bibnamefont {Mosina}}, \bibinfo
  {author} {\bibfnamefont {A.}~\bibnamefont {Papavasileiou}}, \bibinfo {author}
  {\bibfnamefont {K.}~\bibnamefont {Watanabe}}, \bibinfo {author}
  {\bibfnamefont {T.}~\bibnamefont {Taniguchi}}, \bibinfo {author}
  {\bibfnamefont {P.}~\bibnamefont {Maletinsky}}, \bibinfo {author}
  {\bibfnamefont {M.~M.}\ \bibnamefont {Glazov}}, \bibinfo {author}
  {\bibfnamefont {Z.}~\bibnamefont {Sofer}}, \emph {et~al.},\ }\bibfield
  {title} {\bibinfo {title} {Doping-control of excitons and magnetism in
  few-layer {CrSBr}},\ } {\bibfield  {journal} {\bibinfo  {journal}
  {Nature Communications}\ }\textbf {\bibinfo {volume} {15}},\ \bibinfo {pages}
  {4735} (\bibinfo {year} {2024})}\BibitemShut {NoStop}%
\bibitem [{\citenamefont {Graham}\ \emph {et~al.}(2026)\citenamefont {Graham},
  \citenamefont {Wang}, \citenamefont {Nair}, \citenamefont {Mosina},
  \citenamefont {Watanabe}, \citenamefont {Taniguchi}, \citenamefont {Sofer},\
  and\ \citenamefont {Zhou}}]{graham2026space}%
  \BibitemOpen
  \bibfield  {author} {\bibinfo {author} {\bibfnamefont {T.~K.}\ \bibnamefont
  {Graham}}, \bibinfo {author} {\bibfnamefont {Y.-X.}\ \bibnamefont {Wang}},
  \bibinfo {author} {\bibfnamefont {N.~R.}\ \bibnamefont {Nair}}, \bibinfo
  {author} {\bibfnamefont {K.}~\bibnamefont {Mosina}}, \bibinfo {author}
  {\bibfnamefont {K.}~\bibnamefont {Watanabe}}, \bibinfo {author}
  {\bibfnamefont {T.}~\bibnamefont {Taniguchi}}, \bibinfo {author}
  {\bibfnamefont {Z.}~\bibnamefont {Sofer}},\ and\ \bibinfo {author}
  {\bibfnamefont {B.~B.}\ \bibnamefont {Zhou}},\ }\bibfield  {title} {\bibinfo
  {title} {Space-charge-limited van der waals spin transistor},\ }
  {\bibfield  {journal} {\bibinfo  {journal} {Physical Review Letters}\
  }\textbf {\bibinfo {volume} {137}},\ \bibinfo {pages} {040802} (\bibinfo
  {year} {2026})}\BibitemShut {NoStop}%
\bibitem [{\citenamefont {Mondal}\ \emph {et~al.}(2026)\citenamefont {Mondal},
  \citenamefont {Verma}, \citenamefont {Lan}, \citenamefont {Krelle},
  \citenamefont {Hopf}, \citenamefont {Tan}, \citenamefont {von Klitzing},
  \citenamefont {Watanabe}, \citenamefont {Taniguchi}, \citenamefont {Mosina}
  \emph {et~al.}}]{mondal2026twist}%
  \BibitemOpen
  \bibfield  {author} {\bibinfo {author} {\bibfnamefont {P.}~\bibnamefont
  {Mondal}}, \bibinfo {author} {\bibfnamefont {S.}~\bibnamefont {Verma}},
  \bibinfo {author} {\bibfnamefont {W.}~\bibnamefont {Lan}}, \bibinfo {author}
  {\bibfnamefont {L.}~\bibnamefont {Krelle}}, \bibinfo {author} {\bibfnamefont
  {L.}~\bibnamefont {Hopf}}, \bibinfo {author} {\bibfnamefont {R.}~\bibnamefont
  {Tan}}, \bibinfo {author} {\bibfnamefont {R.}~\bibnamefont {von Klitzing}},
  \bibinfo {author} {\bibfnamefont {K.}~\bibnamefont {Watanabe}}, \bibinfo
  {author} {\bibfnamefont {T.}~\bibnamefont {Taniguchi}}, \bibinfo {author}
  {\bibfnamefont {K.}~\bibnamefont {Mosina}}, \emph {et~al.},\ }\bibfield
  {title} {\bibinfo {title} {Twist-tuned exchange and hysteresis in a bilayer
  van der waals magnet},\ } {\bibfield  {journal} {\bibinfo
  {journal} {Nature Communications}\ }\textbf {\bibinfo {volume} {17}},\
  \bibinfo {pages} {5984} (\bibinfo {year} {2026})}\BibitemShut {NoStop}%
\bibitem [{\citenamefont {Chen}\ \emph {et~al.}(2024)\citenamefont {Chen},
  \citenamefont {Samanta}, \citenamefont {Shahed}, \citenamefont {Zhang},
  \citenamefont {Fang}, \citenamefont {Ernst}, \citenamefont {Tsymbal},\ and\
  \citenamefont {Parkin}}]{chen2024twist}%
  \BibitemOpen
  \bibfield  {author} {\bibinfo {author} {\bibfnamefont {Y.}~\bibnamefont
  {Chen}}, \bibinfo {author} {\bibfnamefont {K.}~\bibnamefont {Samanta}},
  \bibinfo {author} {\bibfnamefont {N.~A.}\ \bibnamefont {Shahed}}, \bibinfo
  {author} {\bibfnamefont {H.}~\bibnamefont {Zhang}}, \bibinfo {author}
  {\bibfnamefont {C.}~\bibnamefont {Fang}}, \bibinfo {author} {\bibfnamefont
  {A.}~\bibnamefont {Ernst}}, \bibinfo {author} {\bibfnamefont {E.~Y.}\
  \bibnamefont {Tsymbal}},\ and\ \bibinfo {author} {\bibfnamefont {S.~S.}\
  \bibnamefont {Parkin}},\ }\bibfield  {title} {\bibinfo {title}
  {Twist-assisted all-antiferromagnetic tunnel junction in the atomic limit},\
  } {\bibfield  {journal} {\bibinfo  {journal} {Nature}\ }\textbf
  {\bibinfo {volume} {632}},\ \bibinfo {pages} {1045} (\bibinfo {year}
  {2024})}\BibitemShut {NoStop}%
\bibitem [{\citenamefont {Badola}\ \emph {et~al.}(2026)\citenamefont {Badola},
  \citenamefont {Pawbake}, \citenamefont {Wu}, \citenamefont {S{\"o}ll},
  \citenamefont {Sofer}, \citenamefont {Heid},\ and\ \citenamefont
  {Faugeras}}]{badola2026van}%
  \BibitemOpen
  \bibfield  {author} {\bibinfo {author} {\bibfnamefont {S.}~\bibnamefont
  {Badola}}, \bibinfo {author} {\bibfnamefont {A.}~\bibnamefont {Pawbake}},
  \bibinfo {author} {\bibfnamefont {B.}~\bibnamefont {Wu}}, \bibinfo {author}
  {\bibfnamefont {A.}~\bibnamefont {S{\"o}ll}}, \bibinfo {author} {\bibfnamefont
  {Z.}~\bibnamefont {Sofer}}, \bibinfo {author} {\bibfnamefont
  {R.}~\bibnamefont {Heid}},\ and\ \bibinfo {author} {\bibfnamefont
  {C.}~\bibnamefont {Faugeras}},\ }\bibfield  {title} {\bibinfo {title} {Van
  der waals {CrSBr} alloys with tunable magnetic and optical properties},\
  } {\bibfield  {journal} {\bibinfo  {journal} {Nano Letters}\ }
  (\bibinfo {year} {2026})}\BibitemShut {NoStop}%
\bibitem [{\citenamefont {Smiertka}\ \emph {et~al.}(2026)\citenamefont
  {Smiertka}, \citenamefont {Janikowska}, \citenamefont {Olkowska-Pucko},
  \citenamefont {Krasucki}, \citenamefont {Posmyk}, \citenamefont {Peksa},
  \citenamefont {Surrente}, \citenamefont {Pashov}, \citenamefont {Mosina},
  \citenamefont {Sofer} \emph {et~al.}}]{smiertka2026tunable}%
  \BibitemOpen
  \bibfield  {author} {\bibinfo {author} {\bibfnamefont {M.}~\bibnamefont
  {Smiertka}}, \bibinfo {author} {\bibfnamefont {O.}~\bibnamefont
  {Janikowska}}, \bibinfo {author} {\bibfnamefont {K.}~\bibnamefont
  {Olkowska-Pucko}}, \bibinfo {author} {\bibfnamefont {G.}~\bibnamefont
  {Krasucki}}, \bibinfo {author} {\bibfnamefont {K.}~\bibnamefont {Posmyk}},
  \bibinfo {author} {\bibfnamefont {P.}~\bibnamefont {Peksa}}, \bibinfo
  {author} {\bibfnamefont {A.}~\bibnamefont {Surrente}}, \bibinfo {author}
  {\bibfnamefont {D.}~\bibnamefont {Pashov}}, \bibinfo {author} {\bibfnamefont
  {K.}~\bibnamefont {Mosina}}, \bibinfo {author} {\bibfnamefont
  {Z.}~\bibnamefont {Sofer}}, \emph {et~al.},\ }\bibfield  {title} {\bibinfo
  {title} {Tunable magneto-excitonic coupling in alloyed van der waals
  antiferromagnet},\ } {\bibfield  {journal} {\bibinfo  {journal}
  {arXiv preprint arXiv:2607.14723}\ } (\bibinfo {year} {2026})}\BibitemShut
  {NoStop}%
\bibitem [{\citenamefont {Long}\ \emph {et~al.}(2023)\citenamefont {Long},
  \citenamefont {Ghorbani-Asl}, \citenamefont {Mosina}, \citenamefont {Li},
  \citenamefont {Lin}, \citenamefont {Ganss}, \citenamefont {H\"ubner},
  \citenamefont {Sofer}, \citenamefont {Dirnberger}, \citenamefont {Kamra}
  \emph {et~al.}}]{long2023ferromagnetic}%
  \BibitemOpen
  \bibfield  {author} {\bibinfo {author} {\bibfnamefont {F.}~\bibnamefont
  {Long}}, \bibinfo {author} {\bibfnamefont {M.}~\bibnamefont {Ghorbani-Asl}},
  \bibinfo {author} {\bibfnamefont {K.}~\bibnamefont {Mosina}}, \bibinfo
  {author} {\bibfnamefont {Y.}~\bibnamefont {Li}}, \bibinfo {author}
  {\bibfnamefont {K.}~\bibnamefont {Lin}}, \bibinfo {author} {\bibfnamefont
  {F.}~\bibnamefont {Ganss}}, \bibinfo {author} {\bibfnamefont
  {R.}~\bibnamefont {H\"ubner}}, \bibinfo {author} {\bibfnamefont
  {Z.}~\bibnamefont {Sofer}}, \bibinfo {author} {\bibfnamefont
  {F.}~\bibnamefont {Dirnberger}}, \bibinfo {author} {\bibfnamefont
  {A.}~\bibnamefont {Kamra}}, \emph {et~al.},\ }\bibfield  {title} {\bibinfo
  {title} {Ferromagnetic interlayer coupling in {CrSBr} crystals irradiated by
  ions},\ } {\bibfield  {journal} {\bibinfo  {journal} {Nano
  Letters}\ }\textbf {\bibinfo {volume} {23}},\ \bibinfo {pages} {8468}
  (\bibinfo {year} {2023})}\BibitemShut {NoStop}%
\bibitem [{\citenamefont {Long}\ \emph {et~al.}(2024)\citenamefont {Long},
  \citenamefont {Li}, \citenamefont {Cheng}, \citenamefont {Mosina},
  \citenamefont {Kentsch}, \citenamefont {Sofer}, \citenamefont {Prucnal},
  \citenamefont {Helm},\ and\ \citenamefont {Zhou}}]{long2024rise}%
  \BibitemOpen
  \bibfield  {author} {\bibinfo {author} {\bibfnamefont {F.}~\bibnamefont
  {Long}}, \bibinfo {author} {\bibfnamefont {Y.}~\bibnamefont {Li}}, \bibinfo
  {author} {\bibfnamefont {Y.}~\bibnamefont {Cheng}}, \bibinfo {author}
  {\bibfnamefont {K.}~\bibnamefont {Mosina}}, \bibinfo {author} {\bibfnamefont
  {U.}~\bibnamefont {Kentsch}}, \bibinfo {author} {\bibfnamefont
  {Z.}~\bibnamefont {Sofer}}, \bibinfo {author} {\bibfnamefont
  {S.}~\bibnamefont {Prucnal}}, \bibinfo {author} {\bibfnamefont
  {M.}~\bibnamefont {Helm}},\ and\ \bibinfo {author} {\bibfnamefont
  {S.}~\bibnamefont {Zhou}},\ }\bibfield  {title} {\bibinfo {title} {Rise and
  fall of the ferromagnetism in {CrSBr} flakes by non-magnetic ion irradiation},\
  } {\bibfield  {journal} {\bibinfo  {journal} {Advanced Physics
  Research}\ }\textbf {\bibinfo {volume} {3}},\ \bibinfo {pages} {2400053}
  (\bibinfo {year} {2024})}\BibitemShut {NoStop}%
\bibitem [{\citenamefont {Markina}\ \emph {et~al.}(2026)\citenamefont
  {Markina}, \citenamefont {Pfister}, \citenamefont {Mondal}, \citenamefont
  {Krelle}, \citenamefont {Shradha}, \citenamefont {Von~Klitzing},
  \citenamefont {Mosina}, \citenamefont {Sofer}, \citenamefont {Long},
  \citenamefont {Kentsch}, \citenamefont {Zhou},\ and\ \citenamefont
  {Urbaszek}}]{markina2026detecting}%
  \BibitemOpen
  \bibfield  {author} {\bibinfo {author} {\bibfnamefont {D.~I.}\ \bibnamefont
  {Markina}}, \bibinfo {author} {\bibfnamefont {A.}~\bibnamefont {Pfister}},
  \bibinfo {author} {\bibfnamefont {P.}~\bibnamefont {Mondal}}, \bibinfo
  {author} {\bibfnamefont {L.}~\bibnamefont {Krelle}}, \bibinfo {author}
  {\bibfnamefont {S.}~\bibnamefont {Shradha}}, \bibinfo {author} {\bibfnamefont
  {R.}~\bibnamefont {Von~Klitzing}}, \bibinfo {author} {\bibfnamefont
  {K.}~\bibnamefont {Mosina}}, \bibinfo {author} {\bibfnamefont
  {Z.}~\bibnamefont {Sofer}}, \bibinfo {author} {\bibfnamefont
  {F.}~\bibnamefont {Long}}, \bibinfo {author} {\bibfnamefont {U.}~\bibnamefont
  {Kentsch}}, \bibinfo {author} {\bibfnamefont {S.}~\bibnamefont {Zhou}},\ and\
  \bibinfo {author} {\bibfnamefont {B.}~\bibnamefont {Urbaszek}},\ }\bibfield
  {title} {\bibinfo {title} {Detecting magnetic phase transitions in
  ion-irradiated {CrSBr} through resonant raman scattering},\ }
  {\bibfield  {journal} {\bibinfo  {journal} {arXiv preprint arXiv:2608.18909}\
  } (\bibinfo {year} {2026})}\BibitemShut {NoStop}%
\bibitem [{\citenamefont {Lee}\ \emph {et~al.}(2021)\citenamefont {Lee},
  \citenamefont {Dismukes}, \citenamefont {Telford}, \citenamefont {Wiscons},
  \citenamefont {Wang}, \citenamefont {Xu}, \citenamefont {Nuckolls},
  \citenamefont {Dean}, \citenamefont {Roy},\ and\ \citenamefont
  {Zhu}}]{lee2021magnetic}%
  \BibitemOpen
  \bibfield  {author} {\bibinfo {author} {\bibfnamefont {K.}~\bibnamefont
  {Lee}}, \bibinfo {author} {\bibfnamefont {A.~H.}\ \bibnamefont {Dismukes}},
  \bibinfo {author} {\bibfnamefont {E.~J.}\ \bibnamefont {Telford}}, \bibinfo
  {author} {\bibfnamefont {R.~A.}\ \bibnamefont {Wiscons}}, \bibinfo {author}
  {\bibfnamefont {J.}~\bibnamefont {Wang}}, \bibinfo {author} {\bibfnamefont
  {X.}~\bibnamefont {Xu}}, \bibinfo {author} {\bibfnamefont {C.}~\bibnamefont
  {Nuckolls}}, \bibinfo {author} {\bibfnamefont {C.~R.}\ \bibnamefont {Dean}},
  \bibinfo {author} {\bibfnamefont {X.}~\bibnamefont {Roy}},\ and\ \bibinfo
  {author} {\bibfnamefont {X.}~\bibnamefont {Zhu}},\ }\bibfield  {title}
  {\bibinfo {title} {Magnetic order and symmetry in the 2d semiconductor
  {CrSBr}},\ } {\bibfield  {journal} {\bibinfo  {journal} {Nano
  Letters}\ }\textbf {\bibinfo {volume} {21}},\ \bibinfo {pages} {3511}
  (\bibinfo {year} {2021})}\BibitemShut {NoStop}%
\bibitem [{\citenamefont {Tschudin}\ \emph {et~al.}(2024)\citenamefont
  {Tschudin}, \citenamefont {Broadway}, \citenamefont {Siegwolf}, \citenamefont
  {Schrader}, \citenamefont {Telford}, \citenamefont {Gross}, \citenamefont
  {Cox}, \citenamefont {Dubois}, \citenamefont {Chica}, \citenamefont
  {Rama-Eiroa} \emph {et~al.}}]{tschudin2024imaging}%
  \BibitemOpen
  \bibfield  {author} {\bibinfo {author} {\bibfnamefont {M.~A.}\ \bibnamefont
  {Tschudin}}, \bibinfo {author} {\bibfnamefont {D.~A.}\ \bibnamefont
  {Broadway}}, \bibinfo {author} {\bibfnamefont {P.}~\bibnamefont {Siegwolf}},
  \bibinfo {author} {\bibfnamefont {C.}~\bibnamefont {Schrader}}, \bibinfo
  {author} {\bibfnamefont {E.~J.}\ \bibnamefont {Telford}}, \bibinfo {author}
  {\bibfnamefont {B.}~\bibnamefont {Gross}}, \bibinfo {author} {\bibfnamefont
  {J.}~\bibnamefont {Cox}}, \bibinfo {author} {\bibfnamefont {A.~E.}\
  \bibnamefont {Dubois}}, \bibinfo {author} {\bibfnamefont {D.~G.}\
  \bibnamefont {Chica}}, \bibinfo {author} {\bibfnamefont {R.}~\bibnamefont
  {Rama-Eiroa}}, \emph {et~al.},\ }\bibfield  {title} {\bibinfo {title}
  {Imaging nanomagnetism and magnetic phase transitions in atomically thin
  {CrSBr}},\ } {\bibfield  {journal} {\bibinfo  {journal} {Nature
  Communications}\ }\textbf {\bibinfo {volume} {15}},\ \bibinfo {pages} {6005}
  (\bibinfo {year} {2024})}\BibitemShut {NoStop}%
\bibitem [{\citenamefont {Bagani}\ \emph {et~al.}(2024)\citenamefont {Bagani},
  \citenamefont {Vervelaki}, \citenamefont {Jetter}, \citenamefont
  {Devarakonda}, \citenamefont {Tschudin}, \citenamefont {Gross}, \citenamefont
  {Chica}, \citenamefont {Broadway}, \citenamefont {Dean}, \citenamefont {Roy}
  \emph {et~al.}}]{bagani2024imaging}%
  \BibitemOpen
  \bibfield  {author} {\bibinfo {author} {\bibfnamefont {K.}~\bibnamefont
  {Bagani}}, \bibinfo {author} {\bibfnamefont {A.}~\bibnamefont {Vervelaki}},
  \bibinfo {author} {\bibfnamefont {D.}~\bibnamefont {Jetter}}, \bibinfo
  {author} {\bibfnamefont {A.}~\bibnamefont {Devarakonda}}, \bibinfo {author}
  {\bibfnamefont {M.~A.}\ \bibnamefont {Tschudin}}, \bibinfo {author}
  {\bibfnamefont {B.}~\bibnamefont {Gross}}, \bibinfo {author} {\bibfnamefont
  {D.~G.}\ \bibnamefont {Chica}}, \bibinfo {author} {\bibfnamefont {D.~A.}\
  \bibnamefont {Broadway}}, \bibinfo {author} {\bibfnamefont {C.~R.}\
  \bibnamefont {Dean}}, \bibinfo {author} {\bibfnamefont {X.}~\bibnamefont
  {Roy}}, \emph {et~al.},\ }\bibfield  {title} {\bibinfo {title} {Imaging
  strain-controlled magnetic reversal in thin {CrSBr}},\ }
  {\bibfield  {journal} {\bibinfo  {journal} {Nano Letters}\ }\textbf {\bibinfo
  {volume} {24}},\ \bibinfo {pages} {13068} (\bibinfo {year}
  {2024})}\BibitemShut {NoStop}%
\bibitem [{\citenamefont {{\L}opion}\ \emph {et~al.}(2025)\citenamefont
  {{\L}opion}, \citenamefont {Piel}, \citenamefont {Kliewer}, \citenamefont
  {Terbeck}, \citenamefont {Larusch}, \citenamefont {Henz}, \citenamefont
  {Hei{\ss}enb{\"u}ttel}, \citenamefont {Mosina}, \citenamefont {Deilmann},
  \citenamefont {Rohlfing} \emph {et~al.}}]{lopion2025optical}%
  \BibitemOpen
  \bibfield  {author} {\bibinfo {author} {\bibfnamefont {A.}~\bibnamefont
  {{\L}opion}}, \bibinfo {author} {\bibfnamefont {P.-M.}\ \bibnamefont {Piel}},
  \bibinfo {author} {\bibfnamefont {T.}~\bibnamefont {Kliewer}}, \bibinfo
  {author} {\bibfnamefont {M.}~\bibnamefont {Terbeck}}, \bibinfo {author}
  {\bibfnamefont {J.-H.}\ \bibnamefont {Larusch}}, \bibinfo {author}
  {\bibfnamefont {J.}~\bibnamefont {Henz}}, \bibinfo {author} {\bibfnamefont
  {M.-C.}\ \bibnamefont {Hei{\ss}enb{\"u}ttel}}, \bibinfo {author}
  {\bibfnamefont {K.}~\bibnamefont {Mosina}}, \bibinfo {author} {\bibfnamefont
  {T.}~\bibnamefont {Deilmann}}, \bibinfo {author} {\bibfnamefont
  {M.}~\bibnamefont {Rohlfing}}, \emph {et~al.},\ }\bibfield  {title} {\bibinfo
  {title} {Optical readout of reconfigurable layered magnetic domain structure
  in {CrSBr}},\ } {\bibfield  {journal} {\bibinfo  {journal} {arXiv
  preprint arXiv:2512.04887}\ } (\bibinfo {year} {2025})}\BibitemShut {NoStop}%
\bibitem [{\citenamefont {Sun}\ \emph {et~al.}(2025)\citenamefont {Sun},
  \citenamefont {Hong}, \citenamefont {Chen}, \citenamefont {Sheng},
  \citenamefont {Wu}, \citenamefont {Wang}, \citenamefont {Liang},
  \citenamefont {Liu}, \citenamefont {Yuan}, \citenamefont {Wu} \emph
  {et~al.}}]{sun2025resolving}%
  \BibitemOpen
  \bibfield  {author} {\bibinfo {author} {\bibfnamefont {Z.}~\bibnamefont
  {Sun}}, \bibinfo {author} {\bibfnamefont {C.}~\bibnamefont {Hong}}, \bibinfo
  {author} {\bibfnamefont {Y.}~\bibnamefont {Chen}}, \bibinfo {author}
  {\bibfnamefont {Z.}~\bibnamefont {Sheng}}, \bibinfo {author} {\bibfnamefont
  {S.}~\bibnamefont {Wu}}, \bibinfo {author} {\bibfnamefont {Z.}~\bibnamefont
  {Wang}}, \bibinfo {author} {\bibfnamefont {B.}~\bibnamefont {Liang}},
  \bibinfo {author} {\bibfnamefont {W.-T.}\ \bibnamefont {Liu}}, \bibinfo
  {author} {\bibfnamefont {Z.}~\bibnamefont {Yuan}}, \bibinfo {author}
  {\bibfnamefont {Y.}~\bibnamefont {Wu}}, \emph {et~al.},\ }\bibfield  {title}
  {\bibinfo {title} {Resolving and routing magnetic polymorphs in a 2d layered
  antiferromagnet},\ } {\bibfield  {journal} {\bibinfo  {journal}
  {Nature Materials}\ ,\ \bibinfo {pages} {1}} (\bibinfo {year}
  {2025})}\BibitemShut {NoStop}%
\bibitem [{\citenamefont {Shao}\ \emph {et~al.}(2025)\citenamefont {Shao},
  \citenamefont {Dirnberger}, \citenamefont {Qiu}, \citenamefont {Acharya},
  \citenamefont {Terres}, \citenamefont {Telford}, \citenamefont {Pashov},
  \citenamefont {Kim}, \citenamefont {Ruta}, \citenamefont {Chica} \emph
  {et~al.}}]{shao2025magnetically}%
  \BibitemOpen
  \bibfield  {author} {\bibinfo {author} {\bibfnamefont {Y.}~\bibnamefont
  {Shao}}, \bibinfo {author} {\bibfnamefont {F.}~\bibnamefont {Dirnberger}},
  \bibinfo {author} {\bibfnamefont {S.}~\bibnamefont {Qiu}}, \bibinfo {author}
  {\bibfnamefont {S.}~\bibnamefont {Acharya}}, \bibinfo {author} {\bibfnamefont
  {S.}~\bibnamefont {Terres}}, \bibinfo {author} {\bibfnamefont {E.~J.}\
  \bibnamefont {Telford}}, \bibinfo {author} {\bibfnamefont {D.}~\bibnamefont
  {Pashov}}, \bibinfo {author} {\bibfnamefont {B.~S.}\ \bibnamefont {Kim}},
  \bibinfo {author} {\bibfnamefont {F.~L.}\ \bibnamefont {Ruta}}, \bibinfo
  {author} {\bibfnamefont {D.~G.}\ \bibnamefont {Chica}}, \emph {et~al.},\
  }\bibfield  {title} {\bibinfo {title} {Magnetically confined surface and bulk
  excitons in a layered antiferromagnet},\ } {\bibfield  {journal}
  {\bibinfo  {journal} {Nature materials}\ } (\bibinfo {year}
  {2025})}\BibitemShut {NoStop}%
\bibitem [{\citenamefont {Choi}\ \emph {et~al.}(2026)\citenamefont {Choi},
  \citenamefont {Moon}, \citenamefont {Lee}, \citenamefont {Plutnarov{\'a}},
  \citenamefont {Sofer}, \citenamefont {Menon},\ and\ \citenamefont
  {Crooker}}]{choi2026bulk}%
  \BibitemOpen
  \bibfield  {author} {\bibinfo {author} {\bibfnamefont {J.}~\bibnamefont
  {Choi}}, \bibinfo {author} {\bibfnamefont {Y.}~\bibnamefont {Moon}}, \bibinfo
  {author} {\bibfnamefont {D.}~\bibnamefont {Lee}}, \bibinfo {author}
  {\bibfnamefont {I.}~\bibnamefont {Plutnarov{\'a}}}, \bibinfo {author}
  {\bibfnamefont {Z.}~\bibnamefont {Sofer}}, \bibinfo {author} {\bibfnamefont
  {V.~M.}\ \bibnamefont {Menon}},\ and\ \bibinfo {author} {\bibfnamefont
  {S.~A.}\ \bibnamefont {Crooker}},\ }\bibfield  {title} {\bibinfo {title}
  {Bulk and surface excitons in the van der waals magnet {CrSBr}: Magneto-optical
  studies to 55 t},\ } {\bibfield  {journal} {\bibinfo  {journal}
  {Nano Letters}\ } (\bibinfo {year} {2026})}\BibitemShut {NoStop}%
\bibitem [{\citenamefont {{\'S}miertka}\ \emph {et~al.}(2026)\citenamefont
  {{\'S}miertka}, \citenamefont {Ryga{\l}a}, \citenamefont {Posmyk},
  \citenamefont {Peksa}, \citenamefont {Dyksik}, \citenamefont {Pashov},
  \citenamefont {Mosina}, \citenamefont {Sofer}, \citenamefont {van
  Schilfgaarde}, \citenamefont {Dirnberger} \emph
  {et~al.}}]{smiertka2026distinct}%
  \BibitemOpen
  \bibfield  {author} {\bibinfo {author} {\bibfnamefont {M.}~\bibnamefont
  {{\'S}miertka}}, \bibinfo {author} {\bibfnamefont {M.}~\bibnamefont
  {Ryga{\l}a}}, \bibinfo {author} {\bibfnamefont {K.}~\bibnamefont {Posmyk}},
  \bibinfo {author} {\bibfnamefont {P.}~\bibnamefont {Peksa}}, \bibinfo
  {author} {\bibfnamefont {M.}~\bibnamefont {Dyksik}}, \bibinfo {author}
  {\bibfnamefont {D.}~\bibnamefont {Pashov}}, \bibinfo {author} {\bibfnamefont
  {K.}~\bibnamefont {Mosina}}, \bibinfo {author} {\bibfnamefont
  {Z.}~\bibnamefont {Sofer}}, \bibinfo {author} {\bibfnamefont
  {M.}~\bibnamefont {van Schilfgaarde}}, \bibinfo {author} {\bibfnamefont
  {F.}~\bibnamefont {Dirnberger}}, \emph {et~al.},\ }\bibfield  {title}
  {\bibinfo {title} {Distinct magneto-optical response of frenkel and wannier
  excitons in {CrSBr}},\ } {\bibfield  {journal} {\bibinfo
  {journal} {Nature Communications}\ } (\bibinfo {year} {2026})}\BibitemShut
  {NoStop}%
\bibitem [{\citenamefont {Shree}\ \emph {et~al.}(2021)\citenamefont {Shree},
  \citenamefont {Paradisanos}, \citenamefont {Marie}, \citenamefont {Robert},\
  and\ \citenamefont {Urbaszek}}]{shree2021guide}%
  \BibitemOpen
  \bibfield  {author} {\bibinfo {author} {\bibfnamefont {S.}~\bibnamefont
  {Shree}}, \bibinfo {author} {\bibfnamefont {I.}~\bibnamefont {Paradisanos}},
  \bibinfo {author} {\bibfnamefont {X.}~\bibnamefont {Marie}}, \bibinfo
  {author} {\bibfnamefont {C.}~\bibnamefont {Robert}},\ and\ \bibinfo {author}
  {\bibfnamefont {B.}~\bibnamefont {Urbaszek}},\ }\bibfield  {title} {\bibinfo
  {title} {Guide to optical spectroscopy of layered semiconductors},\
  } {\bibfield  {journal} {\bibinfo  {journal} {Nature Reviews
  Physics}\ }\textbf {\bibinfo {volume} {3}},\ \bibinfo {pages} {39} (\bibinfo
  {year} {2021})}\BibitemShut {NoStop}%
\bibitem [{\citenamefont {Komar}\ \emph {et~al.}(2024)\citenamefont {Komar},
  \citenamefont {{\L}opion}, \citenamefont {Goryca}, \citenamefont {Rybak},
  \citenamefont {Wo{\'z}niak}, \citenamefont {Mosina}, \citenamefont
  {S{\"o}ll}, \citenamefont {Sofer}, \citenamefont {Pacuski}, \citenamefont
  {Faugeras} \emph {et~al.}}]{komar2024colossal}%
  \BibitemOpen
  \bibfield  {author} {\bibinfo {author} {\bibfnamefont {R.}~\bibnamefont
  {Komar}}, \bibinfo {author} {\bibfnamefont {A.}~\bibnamefont {{\L}opion}},
  \bibinfo {author} {\bibfnamefont {M.}~\bibnamefont {Goryca}}, \bibinfo
  {author} {\bibfnamefont {M.}~\bibnamefont {Rybak}}, \bibinfo {author}
  {\bibfnamefont {T.}~\bibnamefont {Wo{\'z}niak}}, \bibinfo {author}
  {\bibfnamefont {K.}~\bibnamefont {Mosina}}, \bibinfo {author} {\bibfnamefont
  {A.}~\bibnamefont {S{\"o}ll}}, \bibinfo {author} {\bibfnamefont
  {Z.}~\bibnamefont {Sofer}}, \bibinfo {author} {\bibfnamefont
  {W.}~\bibnamefont {Pacuski}}, \bibinfo {author} {\bibfnamefont
  {C.}~\bibnamefont {Faugeras}}, \emph {et~al.},\ }\bibfield  {title} {\bibinfo
  {title} {Colossal magneto-excitonic effects in 2d van der waals magnetic
  semiconductor {CrSBr}},\ } {\bibfield  {journal} {\bibinfo
  {journal} {arXiv preprint arXiv:2409.00187}\ } (\bibinfo {year}
  {2024})}\BibitemShut {NoStop}%
\bibitem [{\citenamefont {Ziebel}\ \emph {et~al.}(2024)\citenamefont {Ziebel},
  \citenamefont {Feuer}, \citenamefont {Cox}, \citenamefont {Zhu},
  \citenamefont {Dean},\ and\ \citenamefont {Roy}}]{ziebel2024crsbr}%
  \BibitemOpen
  \bibfield  {author} {\bibinfo {author} {\bibfnamefont {M.~E.}\ \bibnamefont
  {Ziebel}}, \bibinfo {author} {\bibfnamefont {M.~L.}\ \bibnamefont {Feuer}},
  \bibinfo {author} {\bibfnamefont {J.}~\bibnamefont {Cox}}, \bibinfo {author}
  {\bibfnamefont {X.}~\bibnamefont {Zhu}}, \bibinfo {author} {\bibfnamefont
  {C.~R.}\ \bibnamefont {Dean}},\ and\ \bibinfo {author} {\bibfnamefont
  {X.}~\bibnamefont {Roy}},\ }\bibfield  {title} {\bibinfo {title} {{CrSBr}: an
  air-stable, two-dimensional magnetic semiconductor},\ }
  {\bibfield  {journal} {\bibinfo  {journal} {Nano Letters}\ }\textbf {\bibinfo
  {volume} {24}},\ \bibinfo {pages} {4319} (\bibinfo {year}
  {2024})}\BibitemShut {NoStop}%
\bibitem [{\citenamefont {Wang}\ \emph {et~al.}(2025)\citenamefont {Wang},
  \citenamefont {Graham}, \citenamefont {Rama-Eiroa}, \citenamefont {Islam},
  \citenamefont {Badarneh}, \citenamefont {Nunes~Gontijo}, \citenamefont
  {Tiwari}, \citenamefont {Adhikari}, \citenamefont {Zhang}, \citenamefont
  {Watanabe} \emph {et~al.}}]{wang2025configurable}%
  \BibitemOpen
  \bibfield  {author} {\bibinfo {author} {\bibfnamefont {Y.-X.}\ \bibnamefont
  {Wang}}, \bibinfo {author} {\bibfnamefont {T.~K.}\ \bibnamefont {Graham}},
  \bibinfo {author} {\bibfnamefont {R.}~\bibnamefont {Rama-Eiroa}}, \bibinfo
  {author} {\bibfnamefont {M.~A.}\ \bibnamefont {Islam}}, \bibinfo {author}
  {\bibfnamefont {M.~H.}\ \bibnamefont {Badarneh}}, \bibinfo {author}
  {\bibfnamefont {R.}~\bibnamefont {Nunes~Gontijo}}, \bibinfo {author}
  {\bibfnamefont {G.~P.}\ \bibnamefont {Tiwari}}, \bibinfo {author}
  {\bibfnamefont {T.}~\bibnamefont {Adhikari}}, \bibinfo {author}
  {\bibfnamefont {X.-Y.}\ \bibnamefont {Zhang}}, \bibinfo {author}
  {\bibfnamefont {K.}~\bibnamefont {Watanabe}}, \emph {et~al.},\ }\bibfield
  {title} {\bibinfo {title} {Configurable antiferromagnetic domains and lateral
  exchange bias in atomically thin crps4},\ } {\bibfield
  {journal} {\bibinfo  {journal} {Nature Materials}\ }\textbf {\bibinfo
  {volume} {24}},\ \bibinfo {pages} {1414} (\bibinfo {year}
  {2025})}\BibitemShut {NoStop}%
\bibitem [{\citenamefont {Pellet-Mary}\ \emph {et~al.}(2025)\citenamefont
  {Pellet-Mary}, \citenamefont {Dutta}, \citenamefont {Tschudin}, \citenamefont
  {Siegwolf}, \citenamefont {Gross}, \citenamefont {Broadway}, \citenamefont
  {Cox}, \citenamefont {Schrader}, \citenamefont {Happacher}, \citenamefont
  {Chica} \emph {et~al.}}]{pellet2025lateral}%
  \BibitemOpen
  \bibfield  {author} {\bibinfo {author} {\bibfnamefont {C.}~\bibnamefont
  {Pellet-Mary}}, \bibinfo {author} {\bibfnamefont {D.}~\bibnamefont {Dutta}},
  \bibinfo {author} {\bibfnamefont {M.~A.}\ \bibnamefont {Tschudin}}, \bibinfo
  {author} {\bibfnamefont {P.}~\bibnamefont {Siegwolf}}, \bibinfo {author}
  {\bibfnamefont {B.}~\bibnamefont {Gross}}, \bibinfo {author} {\bibfnamefont
  {D.~A.}\ \bibnamefont {Broadway}}, \bibinfo {author} {\bibfnamefont
  {J.}~\bibnamefont {Cox}}, \bibinfo {author} {\bibfnamefont {C.}~\bibnamefont
  {Schrader}}, \bibinfo {author} {\bibfnamefont {J.}~\bibnamefont {Happacher}},
  \bibinfo {author} {\bibfnamefont {D.~G.}\ \bibnamefont {Chica}}, \emph
  {et~al.},\ }\bibfield  {title} {\bibinfo {title} {Lateral exchange bias for
  n{\'e}el-vector control in atomically thin antiferromagnets},\ }
  {\bibfield  {journal} {\bibinfo  {journal} {Nature Communications}\ }\textbf
  {\bibinfo {volume} {16}},\ \bibinfo {pages} {9725} (\bibinfo {year}
  {2025})}\BibitemShut {NoStop}%
\bibitem [{\citenamefont {Raja}\ \emph {et~al.}(2019)\citenamefont {Raja},
  \citenamefont {Waldecker}, \citenamefont {Zipfel}, \citenamefont {Cho},
  \citenamefont {Brem}, \citenamefont {Ziegler}, \citenamefont {Kulig},
  \citenamefont {Taniguchi}, \citenamefont {Watanabe}, \citenamefont {Malic}
  \emph {et~al.}}]{raja2019dielectric}%
  \BibitemOpen
  \bibfield  {author} {\bibinfo {author} {\bibfnamefont {A.}~\bibnamefont
  {Raja}}, \bibinfo {author} {\bibfnamefont {L.}~\bibnamefont {Waldecker}},
  \bibinfo {author} {\bibfnamefont {J.}~\bibnamefont {Zipfel}}, \bibinfo
  {author} {\bibfnamefont {Y.}~\bibnamefont {Cho}}, \bibinfo {author}
  {\bibfnamefont {S.}~\bibnamefont {Brem}}, \bibinfo {author} {\bibfnamefont
  {J.~D.}\ \bibnamefont {Ziegler}}, \bibinfo {author} {\bibfnamefont
  {M.}~\bibnamefont {Kulig}}, \bibinfo {author} {\bibfnamefont
  {T.}~\bibnamefont {Taniguchi}}, \bibinfo {author} {\bibfnamefont
  {K.}~\bibnamefont {Watanabe}}, \bibinfo {author} {\bibfnamefont
  {E.}~\bibnamefont {Malic}}, \emph {et~al.},\ }\bibfield  {title} {\bibinfo
  {title} {Dielectric disorder in two-dimensional materials},\ }
  {\bibfield  {journal} {\bibinfo  {journal} {Nature nanotechnology}\ }\textbf
  {\bibinfo {volume} {14}},\ \bibinfo {pages} {832} (\bibinfo {year}
  {2019})}\BibitemShut {NoStop}%
\bibitem [{\citenamefont {Dean}\ \emph {et~al.}(2010)\citenamefont {Dean},
  \citenamefont {Young}, \citenamefont {Meric}, \citenamefont {Lee},
  \citenamefont {Wang}, \citenamefont {Sorgenfrei}, \citenamefont {Watanabe},
  \citenamefont {Taniguchi}, \citenamefont {Kim}, \citenamefont {Shepard} \emph
  {et~al.}}]{dean2010boron}%
  \BibitemOpen
  \bibfield  {author} {\bibinfo {author} {\bibfnamefont {C.~R.}\ \bibnamefont
  {Dean}}, \bibinfo {author} {\bibfnamefont {A.~F.}\ \bibnamefont {Young}},
  \bibinfo {author} {\bibfnamefont {I.}~\bibnamefont {Meric}}, \bibinfo
  {author} {\bibfnamefont {C.}~\bibnamefont {Lee}}, \bibinfo {author}
  {\bibfnamefont {L.}~\bibnamefont {Wang}}, \bibinfo {author} {\bibfnamefont
  {S.}~\bibnamefont {Sorgenfrei}}, \bibinfo {author} {\bibfnamefont
  {K.}~\bibnamefont {Watanabe}}, \bibinfo {author} {\bibfnamefont
  {T.}~\bibnamefont {Taniguchi}}, \bibinfo {author} {\bibfnamefont
  {P.}~\bibnamefont {Kim}}, \bibinfo {author} {\bibfnamefont {K.~L.}\
  \bibnamefont {Shepard}}, \emph {et~al.},\ }\bibfield  {title} {\bibinfo
  {title} {Boron nitride substrates for high-quality graphene electronics},\
  } {\bibfield  {journal} {\bibinfo  {journal} {Nature
  nanotechnology}\ }\textbf {\bibinfo {volume} {5}},\ \bibinfo {pages} {722}
  (\bibinfo {year} {2010})}\BibitemShut {NoStop}%
\bibitem [{\citenamefont {Cadiz}\ \emph {et~al.}(2017)\citenamefont {Cadiz},
  \citenamefont {Courtade}, \citenamefont {Robert}, \citenamefont {Wang},
  \citenamefont {Shen}, \citenamefont {Cai}, \citenamefont {Taniguchi},
  \citenamefont {Watanabe}, \citenamefont {Carrere}, \citenamefont {Lagarde}
  \emph {et~al.}}]{cadiz2017excitonic}%
  \BibitemOpen
  \bibfield  {author} {\bibinfo {author} {\bibfnamefont {F.}~\bibnamefont
  {Cadiz}}, \bibinfo {author} {\bibfnamefont {E.}~\bibnamefont {Courtade}},
  \bibinfo {author} {\bibfnamefont {C.}~\bibnamefont {Robert}}, \bibinfo
  {author} {\bibfnamefont {G.}~\bibnamefont {Wang}}, \bibinfo {author}
  {\bibfnamefont {Y.}~\bibnamefont {Shen}}, \bibinfo {author} {\bibfnamefont
  {H.}~\bibnamefont {Cai}}, \bibinfo {author} {\bibfnamefont {T.}~\bibnamefont
  {Taniguchi}}, \bibinfo {author} {\bibfnamefont {K.}~\bibnamefont {Watanabe}},
  \bibinfo {author} {\bibfnamefont {H.}~\bibnamefont {Carrere}}, \bibinfo
  {author} {\bibfnamefont {D.}~\bibnamefont {Lagarde}}, \emph {et~al.},\
  }\bibfield  {title} {\bibinfo {title} {Excitonic linewidth approaching the
  homogeneous limit in mos 2-based van der waals heterostructures},\
  } {\bibfield  {journal} {\bibinfo  {journal} {Physical Review
  X}\ }\textbf {\bibinfo {volume} {7}},\ \bibinfo {pages} {021026} (\bibinfo
  {year} {2017})}\BibitemShut {NoStop}%
\bibitem [{\citenamefont {Klein}\ \emph {et~al.}(2022)\citenamefont {Klein},
  \citenamefont {Pham}, \citenamefont {Thomsen}, \citenamefont {Curtis},
  \citenamefont {Denneulin}, \citenamefont {Lorke}, \citenamefont {Florian},
  \citenamefont {Steinhoff}, \citenamefont {Wiscons}, \citenamefont {Luxa}
  \emph {et~al.}}]{klein2022control}%
  \BibitemOpen
  \bibfield  {author} {\bibinfo {author} {\bibfnamefont {J.}~\bibnamefont
  {Klein}}, \bibinfo {author} {\bibfnamefont {T.}~\bibnamefont {Pham}},
  \bibinfo {author} {\bibfnamefont {J.}~\bibnamefont {Thomsen}}, \bibinfo
  {author} {\bibfnamefont {J.}~\bibnamefont {Curtis}}, \bibinfo {author}
  {\bibfnamefont {T.}~\bibnamefont {Denneulin}}, \bibinfo {author}
  {\bibfnamefont {M.}~\bibnamefont {Lorke}}, \bibinfo {author} {\bibfnamefont
  {M.}~\bibnamefont {Florian}}, \bibinfo {author} {\bibfnamefont
  {A.}~\bibnamefont {Steinhoff}}, \bibinfo {author} {\bibfnamefont
  {R.}~\bibnamefont {Wiscons}}, \bibinfo {author} {\bibfnamefont
  {J.}~\bibnamefont {Luxa}}, \emph {et~al.},\ }\bibfield  {title} {\bibinfo
  {title} {Control of structure and spin texture in the van der waals layered
  magnet {CrSBr}},\ } {\bibfield  {journal} {\bibinfo  {journal}
  {Nature Communications}\ }\textbf {\bibinfo {volume} {13}},\ \bibinfo {pages}
  {5420} (\bibinfo {year} {2022})}\BibitemShut {NoStop}%
\bibitem [{\citenamefont {Robert}\ \emph {et~al.}(2018)\citenamefont {Robert},
  \citenamefont {Semina}, \citenamefont {Cadiz}, \citenamefont {Manca},
  \citenamefont {Courtade}, \citenamefont {Taniguchi}, \citenamefont
  {Watanabe}, \citenamefont {Cai}, \citenamefont {Tongay}, \citenamefont
  {Lassagne} \emph {et~al.}}]{robert2018optical}%
  \BibitemOpen
  \bibfield  {author} {\bibinfo {author} {\bibfnamefont {C.}~\bibnamefont
  {Robert}}, \bibinfo {author} {\bibfnamefont {M.}~\bibnamefont {Semina}},
  \bibinfo {author} {\bibfnamefont {F.}~\bibnamefont {Cadiz}}, \bibinfo
  {author} {\bibfnamefont {M.}~\bibnamefont {Manca}}, \bibinfo {author}
  {\bibfnamefont {E.}~\bibnamefont {Courtade}}, \bibinfo {author}
  {\bibfnamefont {T.}~\bibnamefont {Taniguchi}}, \bibinfo {author}
  {\bibfnamefont {K.}~\bibnamefont {Watanabe}}, \bibinfo {author}
  {\bibfnamefont {H.}~\bibnamefont {Cai}}, \bibinfo {author} {\bibfnamefont
  {S.}~\bibnamefont {Tongay}}, \bibinfo {author} {\bibfnamefont
  {B.}~\bibnamefont {Lassagne}}, \emph {et~al.},\ }\bibfield  {title} {\bibinfo
  {title} {Optical spectroscopy of excited exciton states in mos 2 monolayers
  in van der waals heterostructures},\ } {\bibfield  {journal}
  {\bibinfo  {journal} {Physical Review Materials}\ }\textbf {\bibinfo {volume}
  {2}},\ \bibinfo {pages} {011001} (\bibinfo {year} {2018})}\BibitemShut
  {NoStop}%
\bibitem [{\citenamefont {Malitson}(1965)}]{malitson1965interspecimen}%
  \BibitemOpen
  \bibfield  {author} {\bibinfo {author} {\bibfnamefont {I.~H.}\ \bibnamefont
  {Malitson}},\ }\bibfield  {title} {\bibinfo {title} {Interspecimen comparison
  of the refractive index of fused silica},\ } {\bibfield
  {journal} {\bibinfo  {journal} {Journal of the optical society of America}\
  }\textbf {\bibinfo {volume} {55}},\ \bibinfo {pages} {1205} (\bibinfo {year}
  {1965})}\BibitemShut {NoStop}%
\bibitem [{\citenamefont {Schinke}\ \emph {et~al.}(2015)\citenamefont
  {Schinke}, \citenamefont {Christian~Peest}, \citenamefont {Schmidt},
  \citenamefont {Brendel}, \citenamefont {Bothe}, \citenamefont {Vogt},
  \citenamefont {Kr{\"o}ger}, \citenamefont {Winter}, \citenamefont
  {Schirmacher}, \citenamefont {Lim} \emph {et~al.}}]{schinke2015uncertainty}%
  \BibitemOpen
  \bibfield  {author} {\bibinfo {author} {\bibfnamefont {C.}~\bibnamefont
  {Schinke}}, \bibinfo {author} {\bibfnamefont {P.}~\bibnamefont
  {Christian~Peest}}, \bibinfo {author} {\bibfnamefont {J.}~\bibnamefont
  {Schmidt}}, \bibinfo {author} {\bibfnamefont {R.}~\bibnamefont {Brendel}},
  \bibinfo {author} {\bibfnamefont {K.}~\bibnamefont {Bothe}}, \bibinfo
  {author} {\bibfnamefont {M.~R.}\ \bibnamefont {Vogt}}, \bibinfo {author}
  {\bibfnamefont {I.}~\bibnamefont {Kr{\"o}ger}}, \bibinfo {author}
  {\bibfnamefont {S.}~\bibnamefont {Winter}}, \bibinfo {author} {\bibfnamefont
  {A.}~\bibnamefont {Schirmacher}}, \bibinfo {author} {\bibfnamefont
  {S.}~\bibnamefont {Lim}}, \emph {et~al.},\ }\bibfield  {title} {\bibinfo
  {title} {Uncertainty analysis for the coefficient of band-to-band absorption
  of crystalline silicon},\ } {\bibfield  {journal} {\bibinfo
  {journal} {Aip Advances}\ }\textbf {\bibinfo {volume} {5}} (\bibinfo {year}
  {2015})}\BibitemShut {NoStop}%
\end{thebibliography}
\end{document}